\documentclass[10pt,preprint2,flushrt]{aastex}

\def\apje{0}           
\ifodd\apje \usepackage{emulateapj5,apjfonts,epsfig} \fi

\newcommand{\rr}{\mbox{\bf r}}
\newcommand{\vv}{\mbox{\bf v}}
\newcommand{\RR}{\mbox{\bf R}}
\newcommand{\VV}{\mbox{\bf V}}
\newcommand{\HI}{\mbox{H\ {\sc i}}}
\newcommand{\N}{\mbox{\it N}}

\begin{document}
 
\title{Tidal Effects in Clusters of Galaxies}
 
\author{Oleg Y. Gnedin}

\affil{Princeton University Observatory, Princeton, NJ~08544\altaffilmark{1}}

\altaffiltext{1}{Present address: Space Telescope Science Institute,
                                  3700 San Martin Drive, Baltimore, MD 21218;
                                  ognedin@stsci.edu}

\begin{abstract}

High-redshift clusters of galaxies show an over-abundance of spirals
by a factor of $2-3$, and the corresponding under-abundance of S0
galaxies, relative to the nearby clusters.  This morphological
evolution can be explained by tidal interactions with neighboring
galaxies and with the hierarchically growing cluster halo.  The
efficiency of tidal interactions depends on the size and structure of
the cluster, as well as on the epoch of its formation.  I simulate the
formation and evolution of Virgo-type clusters in three cosmologies: a
critical density model $\Omega_0=1$, an open model $\Omega_0=0.4$, and
a flat model $\Omega_0=0.4$ with a cosmological constant.  The orbits
of identified halos are traced with a high temporal resolution ($\sim
10^7$ yr).  Halos with low relative velocities merge only shortly
after entering the cluster; after virialization mergers are
suppressed.

The dynamical evolution of galaxies is determined by the tidal field
along their trajectories.  The maxima of the tidal force do not always
correspond to closest approach to the cluster center.  They are
produced to a large extent by the local density structures, such as
the massive galaxies and the unvirialized remnants of infalling groups
of galaxies.  Collisions of galaxies are intensified by the
substructure, with about 10 encounters within 10 kpc per galaxy in the
Hubble time.  These very close encounters add an important amount
(10--50\%) of the total heating rate.  The integrated effect of
tidal interactions is insufficient to transform a spiral galaxy into
an elliptical, but can produce an S0 galaxy.  Overall, tidal heating
is stronger in the low $\Omega_0$ clusters.

\end{abstract}


\section{Introduction}

This work is motivated by the wealth of recent observational
discoveries of the strong morphological, chemical, and dynamical
evolution of galaxies in clusters.  Transformations of luminous
galaxies, the existence of separate massive dark objects, and the
evolution of cluster substructure intensify the need for theoretical
understanding.  Several different phenomena have been proposed as the
``likely'' explanation (see \S\ref{sec:scenarios}).  While all of them
take place in one form or another, it has eluded our understanding
which process is primarily responsible for the evolution.

I argue that tidal interactions with neighboring galaxies and with the
dark cluster halo are the favored scenario.  Tides operate even before
the cluster dynamically relaxes and affect both stars and gas in the
infalling galaxies.  Tidal effects add enough random motion to
transform thin galactic disks into the kinematically hot thick-disk
configurations characteristic of S0 galaxies.  Tidal disruption of
low density galaxies may provide a large fraction of the diffuse
intracluster light.

The relative importance of tides depends on the cosmological context.
In an $\Omega_0 = 1$ universe, gravitating structures assemble late
and the tidal effects are only now starting to play a role in shaping
the galaxies, whereas in an open universe, $\Omega_0 \sim 0.4$,
galaxies have formed early and experienced a substantial
metamorphosis.  The currently fashionable model, a flat universe with
$\Omega_0 = 0.4$ and a cosmological constant, provides the
intermediate case.  In order to incorporate these effects into the
galactic dynamics self-consistently, I perform numerical simulations
of the evolution of a medium-size cluster of galaxies in the three
cosmologies.

The choice of a numerical scheme is very important here, as the
required dynamical range of simulation is huge: from hundred parsecs
to model the galactic dynamics to tens of Mpc to derive the cluster
tidal field correctly.  In order to achieve such a range with the
existing codes, I introduce a new scheme of separating the cluster
tidal field from the galactic model.  First, I run a lower resolution
cosmological simulation that captures the details of cluster formation
and its time-varying potential.  At some early redshift, $z_g = 5$, I
identify the probable galactic halos and follow their trajectories
until the present.  Along each trajectory, I calculate and store the
components of the tidal shear tensor.  Then, I resimulate individual
galaxies with a very high resolution applying the cluster tidal field
as an external perturbation.  The results of the stellar dynamical
galaxy simulations are presented in a companion paper
(\citealt{PaperII}, hereafter Paper II).  No single simulation of a
cluster of galaxies (\S\ref{sec:prev}) has been able to achieve
comparable resolution on the galactic scale.

I describe a careful procedure to extract the tidal field from the
grid PM simulation.  Because of the particle noise, it is necessary to
apply a smoothing filter extending over a range of several grid cells.
Also, large halos contribute significantly to the tidal field around
their centers.  Since we are interested only in the external tidal force
on the galaxy, this self-contribution needs to be removed.  Special care
is taken to evaluate the contribution of the galactic particles and
subtract it from the total tidal tensor.

Also, the resulting tidal field depends on the correct calculation of
the galactic trajectories.  I use two independent methods to follow
the trajectories: in one, the galactic centers of mass are assumed to
be test particles following the overall cluster potential; in the
other, the center of mass is calculated using the densest subset of
particles constituting an actual halo.  The sampling of halos at each
time step provides an excellent time resolution of the tidal field, of
order $10^7$ yr.  Finally, the distribution of halos and their
relative velocities enable us to estimate the contribution of close
encounters to the external tidal force.

In \S\ref{sec:scenarios}, I review the present observational status of
galaxies in clusters at various redshifts and the theoretical
mechanisms connecting pieces of the puzzle.  In \S\ref{sec:method}, I
describe the numerical method used to simulate the self-consistent
evolution of the three clusters starting with the primordial
fluctuations.  In \S\ref{sec:galaxies}, I identify the galactic halos
and calculate their trajectories.  In \S\ref{sec:gal_distr}, I study
the possible mergers and close encounters between galaxies, including
the effects of halo clustering.  Finally in \S\ref{sec:tidal}, I
calculate the tidal field along the galaxy trajectories and use it in
\S\ref{sec:heating} to estimate the amount of tidal heating of
galaxies.

\section{Scenarios of Galactic Evolution}
  \label{sec:scenarios}

The idea that the environment plays a role in the evolution of
galaxies has been recognized long ago.  \cite{SB:51} suggested that
collisions between galaxies might be responsible for transforming
spirals into the gas-free S0s.  \cite{T:77} advocated a merger of two
spirals to produce an elliptical.  \cite{R:76} argued for the
truncation of dark matter halos of galaxies by the tidal field of the
cluster.  \cite{GG:72} proposed the stripping of gas by the ram
pressure of the intracluster medium.  Finally, \cite{M:83} and most
recently Moore at al. (1998,1999) considered tidal interactions
between the galaxies and the dense cluster core and between the
galaxies themselves.  I review the observational picture and then
discuss briefly the pros and cons of each of these scenarios.

\subsection{``Spirals vs Ellipticals'' or ``Spirals vs S0s''?}
  \label{sec:obs}

The Hubble Space Telescope has been instrumental in shaping our
current understanding of distant clusters of galaxies.  The unique
resolving power of the HST brings us exquisite details of the
distorted profiles of the otherwise normal disk galaxies.  Strongly
depleted in nearby clusters, spirals make up to 50\% of the galaxies
in clusters at redshift $z \sim 0.5$
\citep{Oetal:97,Eetal:97,Setal:97}.  These results are similar to the
\cite{BO:78} effect, an abundance of blue irregular galaxies at an
intermediate redshift.  Spectroscopic evidence \citep{Detal:97}
suggests bursts of recent, or even current, star formation.  This
picture is drastically different from the rich present-day clusters,
populated mostly by featureless elliptical or S0 galaxies which follow
the local morphology-density relation \citep{D:80}.  It is well
established now that a strong chemical and morphological evolution has
taken place in clusters of galaxies.

What are the options?  Did the $z=0.5$ spirals become the $z=0$
ellipticals?  On the contrary, the HST data show that the fraction of
elliptical galaxies {\it increases} with redshift, although not as
strongly as that of spirals.  Both of these effects occur at the
expense of S0s.  Their population is depleted by a factor of two or
three at $z \sim 0.5$ \citep{Detal:97}.  Observations suggest that
star formation has ceased in the ellipticals by $z \sim 0.6$, but
continued in the S0s at least until $z \sim 0.3$ \citep{vDetal:98}.
The ellipticals must have exhausted their supply of fresh gas at an
earlier time, as indicated also by their very high central densities
\citep{K:89}.

Even at low redshift the situation is far from being smooth
\citep{CG:99}.  New ground-based imaging of the Coma and five other
nearby clusters reveals a variety of distorted, tailed, ``line'', and
dwarf galaxies.  These are indicators of the ongoing dynamical
evolution.  Many galaxies are disturbed and some show blue colors.
Also, individual groups of dwarf galaxies have been found in Coma.
These results were not expected for the smooth and regular clusters.

A new observational puzzle has been presented by the discovery of
massive dark objects in the cluster CL 0024+1654, through the
technique of image reconstruction of gravitational lensing
\citep{Tyson:98a,Tyson:98b}.  The dark objects are not associated with
any of the luminous galaxies in the cluster and may contain as much
mass as $10^{10}\, M_{\sun}$.  One of the possible explanations is
that these objects are just ``failed'' galaxies, in which gaseous
dissipation was insufficient to form stars.  Alternatively, they can
be stripped dark matter halos of other galaxies.  Analytical arguments
and previous studies (\S\ref{sec:striphalos}) suggest that extended
galactic halos may be effectively swept away by the tidal field of the
cluster.

\subsection{Mergers of Galaxies}

Collisions of galaxies are assisted by the dense environment of the
cluster.  However, in a virialized cluster, galaxies move with the
relative velocities of the order 1000 km s$^{-1}$ and even on a
head-on trajectory will pass through each other without doing much
harm to the stellar disks.  The gas can be shocked and stripped away
in such a collision \citep{SB:51}, which would leave two gas-free
galaxies.  But a simple estimate argues that close encounters are very
infrequent (however see \S\ref{sec:gal_distr} and Appendix
\ref{sec:app_merger}).

Consider a virialized cluster of galaxies with a one-dimensional
velocity dispersion $\sigma_{cl} = 1000$ km s$^{-1}$ and a virial
radius $R_{cl} = 1$ Mpc.  Assuming that the cluster contains $N_g =
1000$ galaxies uniformly distributed inside this radius, the number of
close encounters a given galaxy of size $R_g = 10$ kpc will experience
in a Hubble time ($t_H = 10^{10}$ yr) is
\begin{equation}
N_{\rm enc,1} \approx {N_g \over \frac{4\pi}{3} R_{cl}^3} \, \pi R_g^2
               \, \sqrt{2} \sigma_{cl} \, t_H  \approx 1.
  \label{eq:enc}
\end{equation}
Here I have neglected gravitational focusing and assumed the relative
velocity to be $\sqrt{2} \, \sigma_{cl}$.  Thus a galaxy is expected
to encounter one other galaxy over the course of its evolution.
However, the probability of a merger of two galaxies is much smaller:
\begin{equation}
P_{\rm mer} \sim N_{\rm enc,1} \, \left( {\sigma_g \over \sigma_{cl}} \right)^4
        \approx 10^{-3}
  \label{eq:mer}
\end{equation}
for the galactic velocity dispersion $\sigma_g = 200\ {\rm km\
s}^{-1}$.  Here, three powers of $\sigma_g/\sigma_{cl}$ come from the
phase space density and another power of $\sigma_g/\sigma_{cl}$ comes
from the limitation of the velocity that leads to a merger
\citep{BT:87}.

Also, ellipticals usually have much higher central densities than
spiral galaxies.  So if two spirals merge, it seems hard to close the
gap in density without strong dissipation \citep{C:86}.  As Jim Gunn
once said \citep{G:87}, ``I do not think you can make rocks by merging
clouds.''  Thus the merger scenario appears unlikely to dominate the
dynamics of galaxies, although mergers could be important for the
formation of the central giant cD galaxy \citep{OH:77,M:85,L:88} and
the ultraluminous infrared galaxies and AGNs (e.g., \citealt{CS:01}).

\subsection{Truncation of Dark Matter Halos}
  \label{sec:striphalos}

\cite{R:76}, \cite{W:76}, and \cite{M:83}, among others, have
investigated the truncation of dark galactic halos by the tidal field
of the cluster.  In the simplest case, where both the cluster and the
galaxy halos have singular isothermal distributions, the tidal radius
is approximately
\begin{equation}
R_t \approx {\sigma_g \over \sigma_{cl}} \, R_p,
\end{equation}
where $R_p$ is the distance of closest approach to the center of the
cluster.

Real clusters are not perfectly smooth.  Any substructure left after
the hierarchical formation of the cluster will amplify the truncation
effect.  It is expected to control the amount of dark matter, and
therefore, the mass of the galaxies.  On the other hand, tidal
truncation should not significantly affect the dynamics of the
galactic disks which are much more compact.  Morphological evolution
of luminous galaxies must be due to something else.

\subsection{Ram-Pressure Stripping of Gas}

\cite{GG:72} argued that the ram pressure of the intracluster medium
(ICM) might sweep away gas from the galactic disks.  X-ray
observations show the presence of a significant amount of hot gas at
the virial temperature, contributing as much as 10\% of the total
cluster mass.  While the stars move freely through this media, the gas
is subject to the ram pressure $P_{\rm ram} \approx \rho_{\rm icm}
v^2$, where $\rho_{\rm icm}$ is the density of the ICM and $v \sim
\sigma_{cl}$ is the velocity of the galaxy.  In order to strip the
gas, the ram pressure should exceed the restoring gravitational force
per unit area of the disk:
\begin{equation}
\rho_{\rm icm} v^2 > 2\pi G \, \Sigma_s \, \Sigma_g,
  \label{eq:ram}
\end{equation}
where $\Sigma_s$ is the surface density of the stars, and $\Sigma_g$
is the surface density of the gas.  For example, if the gas with an
average density 1 cm$^{-3}$ is distributed over a scale height of 100
pc and the stellar density is typical of the solar neighborhood,
$\Sigma_s \approx 75\ M_{\sun}$ pc$^{-2}$, the ICM needs to be as
dense as $n_{\rm icm} > 4\times 10^{-4}$ cm$^{-3}$.  Such densities
are observed in the cores of rich clusters, and numerical experiments
\citep{Ketal:93,AMB:99,QMB:00} show that a typical spiral there may
lose most of its atomic gas.

Spiral galaxies in clusters have long been known to be deficient of
atomic hydrogen \citep{KS:76,HGC:84}.  These ``anemic'' galaxies
\citep{vdB:91} have normal molecular hydrogen abundances but lack the
atomic component in the outer parts of their disks.  Since the H$_2$
gas resides in dense molecular clouds, it is more able to withstand
the external pressure; on the other hand, the \HI\ gas is usually
distributed much more smoothly at lower densities and is susceptible
to the ram pressure stripping.  But \cite{VJ:91} noticed an
interesting feature in three rich Abell clusters and the Virgo
cluster: the \HI\ deficiency is stronger for larger galaxies.  Since
bigger spirals (excluding the LSBs) usually have a higher surface
density, they should be able to protect their gaseous component
better.  The observed trend is the opposite.  Having considered
various possible mechanisms of gas removal, Valluri and Jog conclude
that the \HI\ deficiency can only be explained by tidal interactions.

\subsection{Tidal Interactions}
  \label{sec:prev}

Tides {\it are} a critical factor of the cluster environment.  Tidal
effects depend on relative sizes of the system and the perturber.
Consider stellar systems from smallest to largest scales: stars are
$10^8$ times smaller than star clusters and experience almost no tidal
effects (except in close binaries).  Star clusters are $10^4$ times
smaller than galaxies and experience significant tidal effects, which
however operate on a long time scale \citep{GO:97}.  In contrast,
galaxies are just 30 to 100 times smaller than clusters of galaxies.
There tidal effects ought to be important!

Semi-analytic models of galactic encounters \citep{M:83} predict the
segregation of galaxies by mass and the formation of a central cD
galaxy.  The giant cD grows by swallowing other massive galaxies in a
process that widens the gap between the first and the second ranking
galaxies \citep{OH:77}.  Mass segregation in clusters is also
supported by numerical simulations \citep{Fetal:96}.  Recently,
\cite{D:98} confirmed that the central galaxy, rising by mergers of
its neighbors, follows a de Vaucouleurs profile and would be
classified as a giant elliptical.

Restricted numerical simulations of \cite{BV:90} and \cite{V:93} show
that tidal compression by the cluster halo produces spiral arms and
tidal tails in some of the representative models of disk galaxies.

More recently, Moore at al. (1996b,1998) simulated the dynamics of
galaxies in a fixed, isothermal cluster.  They found that fast close
encounters with massive galaxies tend to destroy many dwarfs, and even
large galaxies lose their thin disks.  So the spiral galaxies are
essentially transformed into the ellipticals or dwarf spheroidals.
However, the potential of a real cluster grows in time, which leads to
two competing effects: (1) galaxies enter the cluster at a later epoch
and have less time to interact (depending on cosmological model), and
(2) the time-varying cluster potential gives rise to large-scale
perturbations in the tidal field and causes tidal shocks comparable to
those from massive galaxies.  New simulations by \cite{Metal:99}
include the hierarchical cluster formation and show weaker evolution
of galaxies.  Whether the galaxy--galaxy or the cluster--galaxy
interactions dominate has not yet been understood.

Similar tidal effects operate on smaller scale in the Local Group and
may lead to the morphological evolution of the dwarf satellite
galaxies \citep{Mayer:01}.

\section{Numerical Method}
  \label{sec:method}

I use a Particle-Mesh code \citep{C:92} to simulate the formation and
evolution of three clusters of galaxies in the critical
($\Omega_0=1$), open ($\Omega_0=0.4$), and flat
($\Omega_0+\Omega_\Lambda=1$) cosmological models described in Table
\ref{tab:cosmo}.  The code solves Poisson's equation using FFT on a
rectangular grid and evaluates the density field using the
Cloud-in-Cell (CIC) method.  The code runs fast in parallel, allowing
a large simulation with $256^3$ particles distributed on a $512^3$
grid.  Some of the PM results have been tested with the Adaptive Mesh
Refinement code \citep{AMR}.  Specifically, I checked galaxy
trajectories and the calculation of the tidal force.

The comoving size of the simulation box is $32 \, h^{-1}$ Mpc and the
grid cell size is $\Delta = 62.5 \, h^{-1}$ kpc.  This is a compromise
between the desire to reach a reasonably high resolution in the
cluster center and the need to have enough computational volume to
include late infall.  The latter is important because a very strong
initial density perturbation may lead to more mass being bound to the
cluster than is available in the whole box, $9.1\times 10^{15}\,
\Omega_0 h^{-1}\, M_{\sun}$, causing the cluster to ``under-form''.

Tests with different amplitudes of the power spectrum, $\sigma_8$,
show that the dark matter velocity dispersion artificially saturates
for $\sigma_8 \gg 1$ in the low $\Omega_0$ models.  Therefore, I adopt
$\sigma_8=1$ for these models, in rough agreement with the relation
derived from the comparison of large scale \N-body simulations with
the all-sky survey of X-ray clusters, $\sigma_8 \approx 0.5\,
\Omega_0^{-0.5}$ \citep{ECF:96,P:98}.  Then the virial mass of the
simulated clusters is about $4\times 10^{14}\, M_{\sun}$, or 7\% of
the total mass within the computational box.  The actual amount of
mass bound to the clusters may be a factor of 2--3 higher, up to 20\%
of the total mass.  In the critical density model, the normalization
$\sigma_8=0.5$ is chosen such that the cluster velocity dispersion is
the same as in the other two models.

The simulations start at redshift $z_i = 20$.  The time step is
determined by the requirement that no particle crosses more than a
half of the grid cell, $0.5 \Delta$.  Also, it limits the expansion
parameter $a \equiv (1+z)^{-1}$ to grow by no more than $0.05
(1+z_i)^{-1}$.  I have checked that increasing the time step by a factor
of two does not change noticeably the structure of the resulting
cluster.

The look-back time as a function of redshift (\citealt{P:93}, p. 314)
can be calculated analytically in each of the three models.  For
$\Omega_0=1$, the standard result is $H_0\, t(z) = {2 \over 3} \,
(1+z)^{-3/2}$.  For $\Omega_\Lambda=0$,
\begin{equation}
H_0\, t(z) = {(1 + \Omega_0 z)^{1/2} \over (1-\Omega_0) (1+z)} +
           {\Omega_0 \over 2 (1-\Omega_0)^{3/2}} \,
            \ln{f^{1/2}-1 \over f^{1/2}+1}, \;\;\;\;
f \equiv {1 + \Omega_0 z \over 1-\Omega_0}.
\end{equation}
For $\Omega_0+\Omega_\Lambda=1$,
\begin{equation}
H_0\, t(z) = {2 \over 3 (1-\Omega_0)^{1/2}} \,
           \ln{1 + (1+f)^{1/2} \over f^{1/2}}, \;\;\;\;
f \equiv {\Omega_0 \over 1-\Omega_0} \, (1+z)^3.
\end{equation}
The Hubble constant is assumed to be $H_0 = 65$ km s$^{-1}$ Mpc$^{-1}$
($h = 0.65$), according to the time delay measurements in a
gravitational lens system 0957+561 \citep{Ketal:97} and other
independent methods \citep{Wendy:94}.

On 16 processors of SGI Origin 2000, the full simulation with a
$512^3$ grid runs for about 36 hours.  A large fraction of the time is
spent on the calculation of the tidal field due to and around the
identified halos.  All simulations presented in this paper were done
using the 64-processor SGI Origin 2000 supercomputer at the Princeton
University Observatory.

\subsection{Constrained Initial Conditions}

The cluster simulations use the constrained Gaussian random initial
conditions, generated using the method of \cite{HR:91}.  The actual
realization is taken from the cluster comparison project of
\cite{Fetal:99}.  In that work the initial density perturbations were
chosen to produce a $3 \sigma$ peak with a Gaussian filter of radius
$5 \, h^{-1}$ Mpc.  Retaining the same random phases from the original
realization, I have renormalized the perturbations according to the
power spectrum of the chosen cosmological models.

For each of the three models, I evaluate the power spectrum $P(k)_{\rm
new}$, including the effects of baryons, using the fitting expressions
of \cite{BBKS:86} and \cite{HS:96}.  On the cluster and sub-cluster
scales, the power spectra follow approximately a power law $P(k)
\propto k^{-2.5}$.  Then I transform the initial displacements $d$
into the Fourier space and renormalize their amplitudes according to
the new power spectrum:
\begin{equation}
d_k^{\rm new} = d_k^{\rm old} \,
                \left({P(k)_{\rm new} \over P(k)_{\rm old}}\right)^{1/2},
\end{equation}
where the old power spectrum, $P(k)_{\rm old}$, is from \cite{Fetal:99}.
Both power spectra are normalized to the corresponding value of
$\sigma_8^2(z_i)$ at the initial redshift $z_i$.  Finally, I transform
the perturbation field back to real space and obtain the initial
displacements for the dark matter particles.

\subsection{Structure of the Clusters}
  \label{sec:clstructure}

The parameters of the resulting clusters are given in Table
\ref{tab:clusters}.  The virialized region is taken to be a sphere
originating at the cluster center, with a mean overdensity $\delta =
200$.  The average mass of the clusters is $4\times 10^{14}\,
M_{\sun}$, and the one-dimensional velocity dispersion is 660 km
s$^{-1}$.

\begin{figure}
\plotone{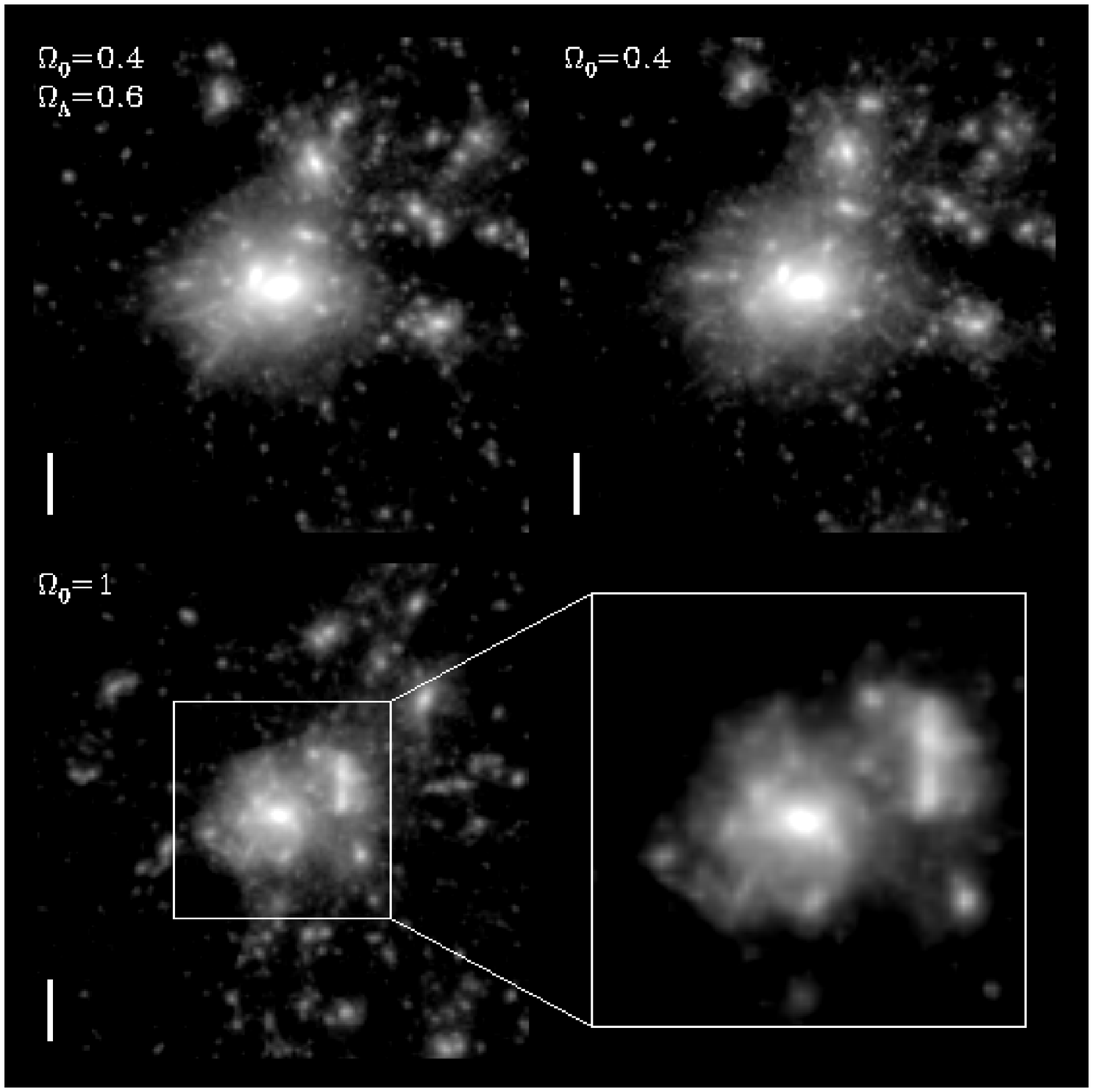}
\caption{The surface density map of the three simulated clusters at
  $z=0$.  The projection is over the $X$ direction.  Saturated white
  regions are denser than 200 times the mean density; regions below the
  mean density are black.  The size of the windows is $8 \, h^{-1}$ Mpc.
  Thick bars in the lower left corner of each panel indicate
  $1 \, h^{-1}$ Mpc.  The lower right panel shows the enlarged view
  of the inner region of the $\Omega_0=1$ cluster.
  \label{fig:den_z0}}
\end{figure}

Figure \ref{fig:den_z0} shows the projected surface density of the
three clusters at the present epoch.  The $\Omega_0=1$ cluster has not
yet relaxed and still continues to accrete matter.  Note a large
infalling group of galaxies to the upper right from the center on the
enlarged panel.  In contrast, the two low $\Omega_0$ clusters appear
more spherical and regular, and also very similar to each other; the
cosmological constant does not seem to affect the cluster properties.

\begin{figure}
\plotone{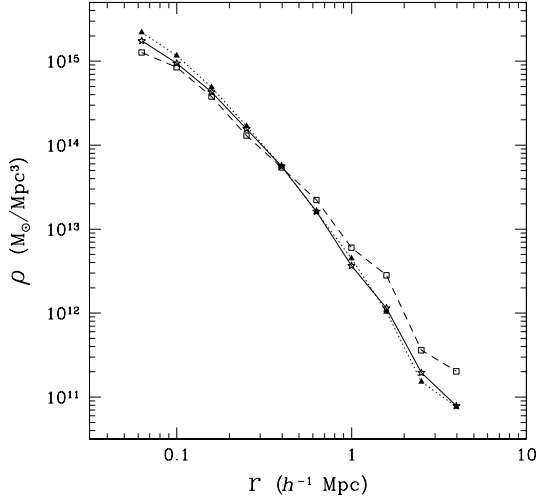}
\caption{The density profile of the three clusters.  Dashes and open squares
  are for the $\Omega_0=1$ model, dots and filled triangles are for the
  $\Omega_0=0.4$ model, solid line and stars are for the $\Omega_0=0.4$,
  $\Omega_\Lambda=0.6$ model.  The innermost point corresponds to the
  resolution element, $\Delta$.
  \label{fig:clden}}
\end{figure}

\begin{figure}
\plotone{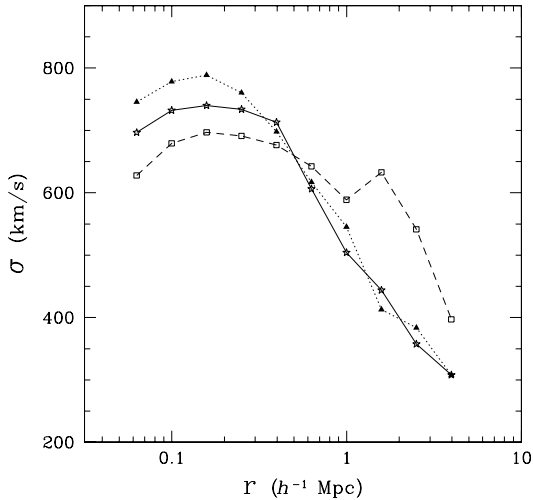}
\caption{The velocity dispersion of dark matter in the three clusters.
  Line notation as in Figure \protect\ref{fig:clden}.  Note a rise of the
  velocity dispersion at 2 $h^{-1}$ Mpc from the center of the
  $\Omega_0=1$ cluster due to the new infalling group of galaxies.
  \label{fig:cldisp}}
\end{figure}

Figure \ref{fig:clden} shows that the density profiles of the three
clusters are very similar, with a slight excursion of the $\Omega_0=1$
cluster at $r \approx 2 \, h^{-1}$ Mpc due to the infalling group.
This group is clearly visible on the velocity dispersion plot, Figure
\ref{fig:cldisp}.

\begin{figure}
\plotone{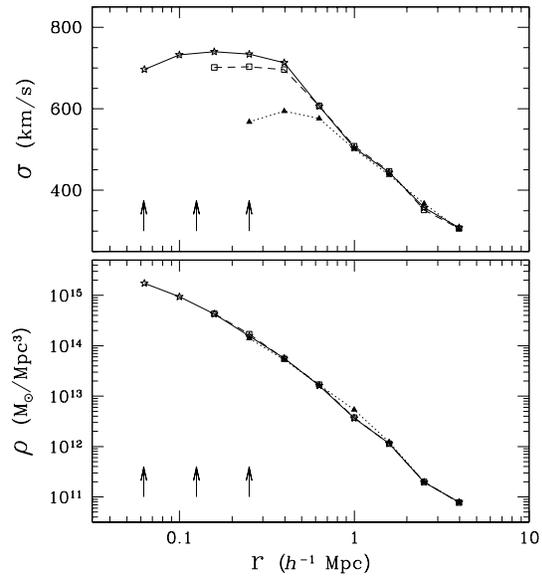}
\caption{Resolution effects on the structure of the $\Omega_0=0.4$,
  $\Omega_\Lambda=0.6$ cluster.  Solid line and stars are for the main
  $512^3$ grid cells simulation, dashes and open squares are for the $256^3$
  simulation, and dots and filled triangles are for the $128^3$
  simulation.  Arrows point at the respective cell sizes, $\Delta$, 
  in each simulation.  The nominal PM force resolution is $2 \Delta$.
  \label{fig:clcomp}}
\end{figure}

The decrease of the velocity dispersion at small distances from the
center is partly real and partly numerical.  The high resolution dark
matter simulations in \cite{Fetal:99} show a similar decrease;
therefore our models should exhibit the trend.  Also, such a decrease
exists in the popular NFW model \citep{NFW:97}.  On the other hand,
present simulations lack resolution in the cluster core and cannot be
trusted at small radii.  To illustrate the resolution effects, I have
repeated the simulations of Cluster III on smaller grids, $256^3$ and
$128^3$ cells (Table \ref{tab:clusters_res}), with the correspondingly
fewer particles, $N_{\rm grid} = 2^3\, N_{\rm p}$.  Figure
\ref{fig:clcomp} shows that the density profile is reproduced well to
the resolution limit of each simulation, but the velocity dispersion
starts to decline much earlier in the lower resolution runs.
Nevertheless, the difference between the $512^3$ and the $256^3$ runs
is small and confirms the accuracy of the larger simulation.

The simulated clusters are similar to the Virgo cluster in size and
velocity dispersion.  ROSAT results \citep{B:94} indicate that Virgo's
X-ray mass is $(1.5-5.5) \times 10^{14}\, M_{\sun}$, while the optical
velocity dispersion of galaxies is between 581 and 721 km s$^{-1}$
\citep{H:85}.  These values are quite common for the nearby clusters,
as shown by the optical and X-ray survey of \cite{Getal:98}.  Often
cited in theoretical work, the Coma cluster is much larger, with the
one-dimensional velocity dispersion about 1000 km s$^{-1}$
\citep{MG:94} and the virial mass of order $2 \times 10^{15}\,
M_{\sun}$ \citep{B:94}.  Our models, therefore, represent modest but
more common clusters of galaxies.  Their velocity dispersion is
characteristic of the Abell richness class 1 (see \citealt{Petal:92}).

\section{Galaxies in the Simulation}
  \label{sec:galaxies}

The definition of a galaxy in the cluster simulation is not trivial.
While at high redshift dark matter halos are significantly denser than
the surrounding media, at low redshift the density contrast
effectively disappears.  This effect is known as overmerging of halos
(e.g., \citealt{MKL:96,Ketal:99}) due to the insufficient mass
resolution.  Therefore, it is difficult and, to some extent, uncertain
to try to identify the galaxy halos from the particle distribution at
all times.  Instead, I have implemented the following simple
procedure, which can be extended to a higher degree of sophistication.

I identify the dark matter halos at a time early enough that the
density contrast is large, but late enough that galaxies are known to
exist.  A fiducial epoch at which the central parts of large galaxies
are assumed to have formed is taken to be $z_g = 5$.

In order to obtain a statistically significant sample, I choose 100
most massive halos identified by the group finding algorithm.  Then, I
add 100 massless particles to the PM simulation with the coordinates
and velocities of the halo centers of mass.  These new ``galactic''
particles do not affect either the density or the potential
calculation but rather act as test particles.  I follow the motion of
these particles and output the tidal field along their trajectories at
each time step.

By not using real particles to calculate galaxy trajectories, we
effectively avoid the overmerging problem: the test particles will not
merge even if they come very close to each other.  The drawback is
that due to a finite time step the test particles may gradually depart
from the bulk of the galactic particles.  To guard against this
effect, I have also implemented another scheme.

The second scheme involves real particles constituting the halo.  I
calculate the halo density at each particle's position using the same
CIC kernel that is used in the main PM calculation.  Then I take the
densest subset of particles, 1/8 of the total, to calculate the center
of mass directly.  These dense particles presumably lie deep in the
potential well of the halo and serve as robust indicators of the
center.  In case of the two galaxies merging, their centers defined
this way would clearly indicate it.  However, I do not attempt to
resolve the internal structure of galaxies in these simulations and
only trace their centers of mass.

\subsection{Group Finding Algorithm}

Dark matter halos are identified using the coordinate-free algorithm
HOP proposed by \cite{EH:98}.  It is similar in spirit to DENMAX
\citep{GB:94} but uses densities at the particle positions instead of
the regular grid.  I have extended HOP by including a procedure for
the removal of particles which are not gravitationally bound to the
group.  This procedure, similar to the SKID algorithm \citep{WHK:97},
is outlined below.

The gravitational energy of the particles in a group is calculated by
direct summation of the Plummer potentials with a smoothing length
corresponding to one grid cell of the cluster simulation.  The kinetic
energy includes the Hubble expansion term.  The total energy per unit
mass is
\begin{equation}
E_i = {1\over 2} \left[\vv_i-\vv_{\rm com} + H_g (\rr_i-\rr_{\rm com})\right]^2
      - \sum_j {G m_p \over ((\rr_i-\rr_j)^2 + \epsilon^2)^{1/2}},
\end{equation}
where $m_p$, $\rr_i$, $\vv_i$ are the particle mass, physical
coordinate, and velocity, respectively, and $H_g$ is the Hubble
constant at redshift $z_g = 5$.  The index ``com'' refers to the
center of mass of the group.  The particles are unbound as follows.  I
find particle $m$ with the maximum energy and discard it, if $E_m >
0$.  I recalculate the center of mass using the remaining particles
and update the potential energy.  Then, I find a new highest energy
particle and repeat the procedure.  In the end, all remaining
particles are bound or else the group is dispersed.

A straightforward implementation of the unbinding procedure works
well, although somewhat slowly for groups containing more than $10^4$
particles.  In the longest case, the unbinding algorithm took about 40
minutes on a single SGI R10000 processor.

The second part of the HOP algorithm, REGROUP, uses three different
overdensity parameters to define the outer boundary of the group
($\delta_{outer}$), the saddle point for merging two groups together
($\delta_{saddle}$), and the minimum density of a viable group
($\delta_{peak}$).  After some experimenting, I have chosen the
following parameters to produce the reasonable-size halos:
[$\delta_{outer},\delta_{saddle},\delta_{peak}$] = [15,45,50] for the
$\Omega_0=1$ model, and [20,55,60] for the low $\Omega_0$ models.  The
different sets of parameters are necessary because of the different
masses of particles ($m_p = 5.4\times 10^8\, \Omega_0 h^{-1}\
M_{\sun}$).

\begin{figure}
\plotone{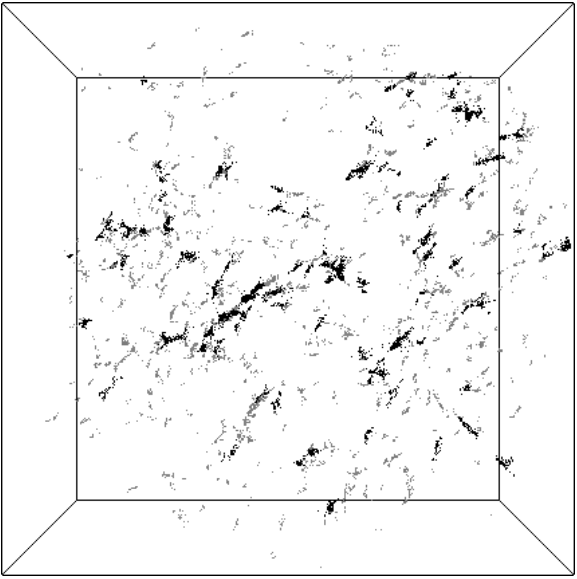}
\caption{Three-dimensional distribution of particles in the identified
  halos at $z_g=5$ in the $\Omega_0=1$ cluster.
  The 100 most massive halos, whose trajectories are followed
  in the simulation, are black; others are gray.  Boundary size is 16
  $h^{-1}$ Mpc comoving.
  \label{fig:gal_om1}}
\end{figure}

\begin{figure}
\plotone{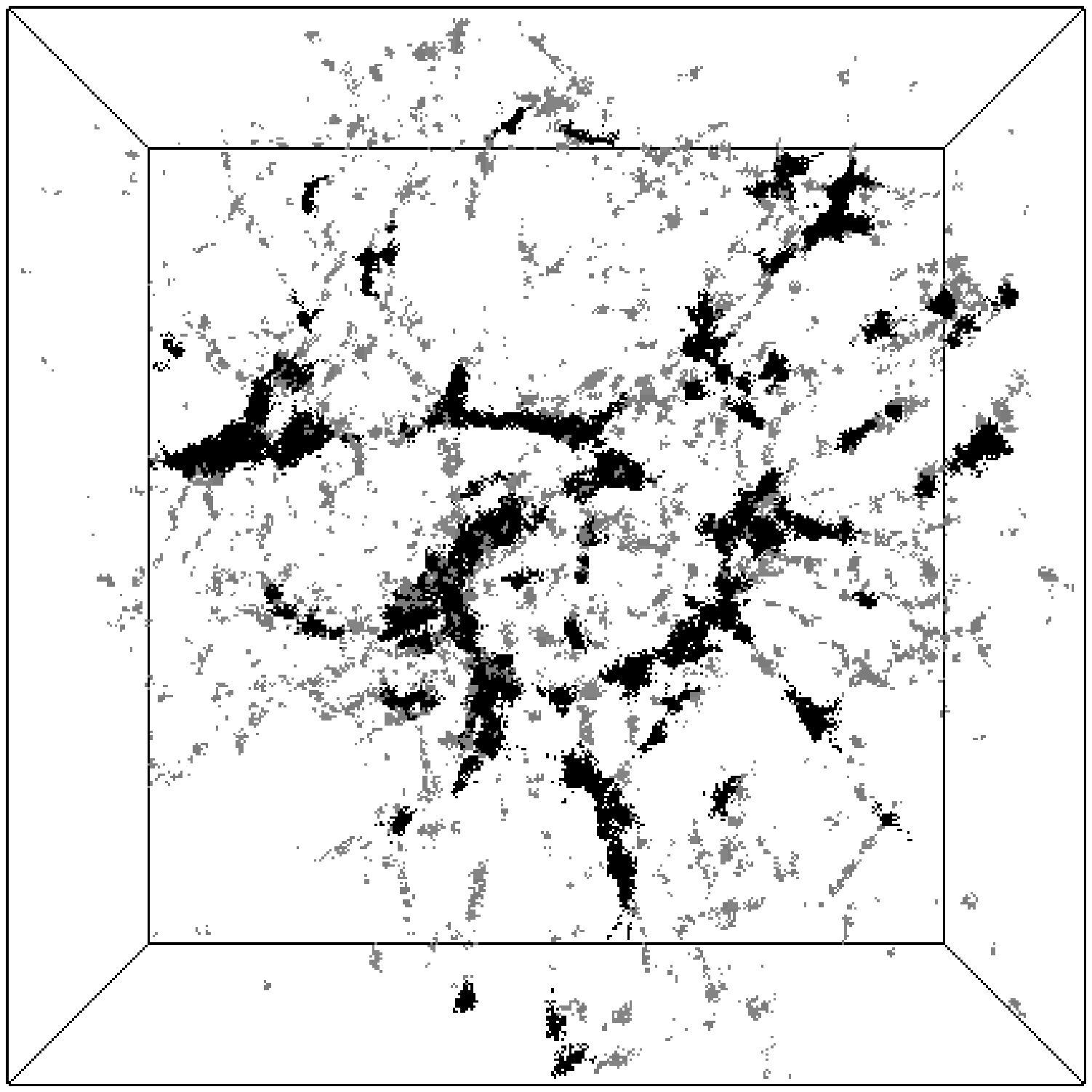}
\caption{Particles in the identified halos in the $\Omega_0=0.4$ cluster.
  \label{fig:gal_om.4}}
\end{figure}

\begin{figure}
\plotone{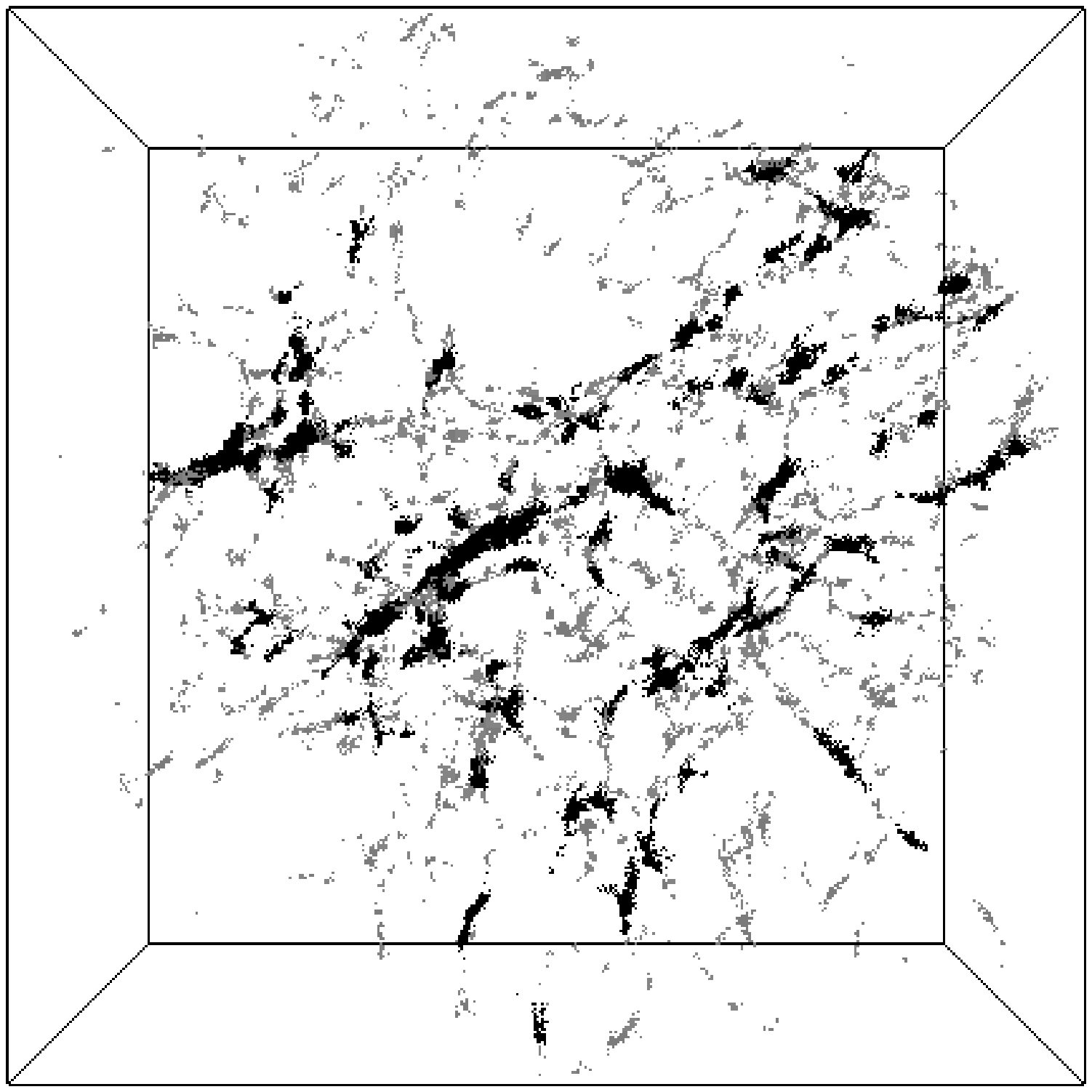}
\caption{Particles in the identified halos in the $\Omega_0=0.4$,
  $\Omega_\Lambda=0.6$ cluster.
  \label{fig:gal_om.4L}}
\end{figure}

Figures \ref{fig:gal_om1}--\ref{fig:gal_om.4L} illustrate the location
of particles in the identified halos.  Notice the filamentary
structure of bound particles indicating the directions along which the
halos continue to assemble.  Most galaxies in the central part of the
simulation volume (shown on the Figures) fall eventually into the
forming cluster.
Because of the difference in particle masses, the $\Omega_0=1$ halos
contain significantly fewer particles than the $\Omega_0=0.4$ halos.

\subsection{Halo Mass Function}
\label{sec:mf}

The mass function of the identified halos can be compared with the
prediction of the Press-Schechter theory \citep{PS:74}.  I have
calculated the expected number density of galaxies at $z_g$ using
equations (A4-A9) of \cite{MMW:98}.

\begin{figure}
\plotone{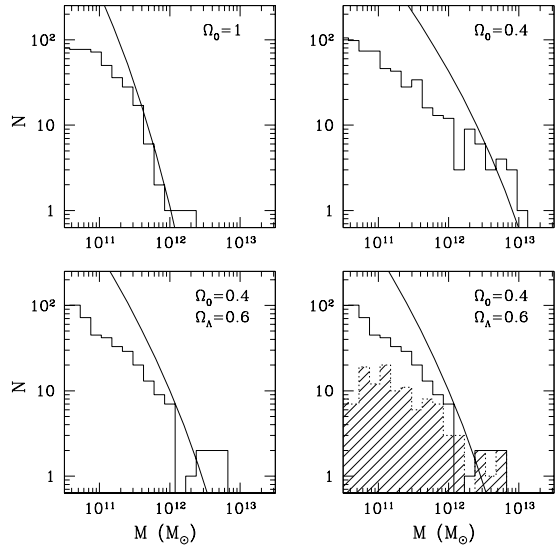}
\caption{The mass function of identified halos in the three
  clusters (histograms).  Superimposed are the predictions of the
  Press-Schechter theory for the corresponding power spectra (lines).
  The lower right panel shows, for comparison, the mass function from
  the lower resolution, $256^3$ simulation of Cluster III (shaded histogram).
  \label{fig:mf_all}}
\end{figure}

Figure \ref{fig:mf_all} shows the halo mass functions in the three
clusters.  Note the relative depletion of low mass halos in the
simulations.  This is partly due to the truncation of the input power
spectrum at the Nyquist frequency, which effectively prevents the
formation of small-scale objects.  For comparison, the lower right
panel shows the mass function obtained in the lower resolution,
$256^3$ simulation of Cluster III.  As expected, the smaller halos are
depleted more strongly than the larger ones.

\subsection{Galaxy Trajectories}
  \label{sec:traject}

\begin{figure}
\plotone{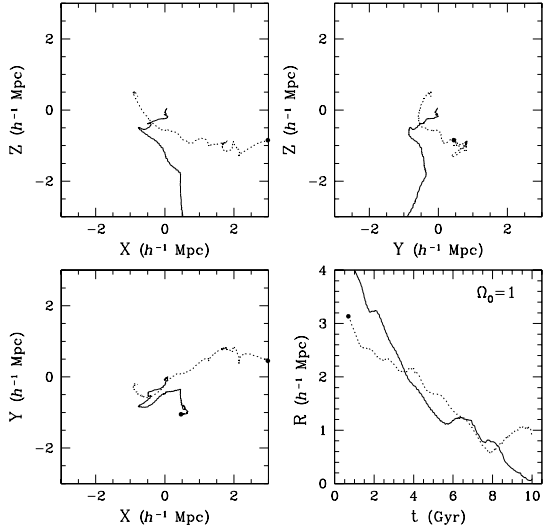}
\caption{Three projections of galaxy trajectories in the
  $\Omega_0=1$ cluster (comoving coordinates).  One galaxy is very
  massive (solid line), ranked \#4 by mass in the list of all
  identified halos, the other is ranked \#93 (dotted line).
  The lower right panel shows the distance from the final cluster center.
  Large dots mark the initial positions where the halos were identified.
  \label{fig:traject_om1}}
\end{figure}

\begin{figure}
\plotone{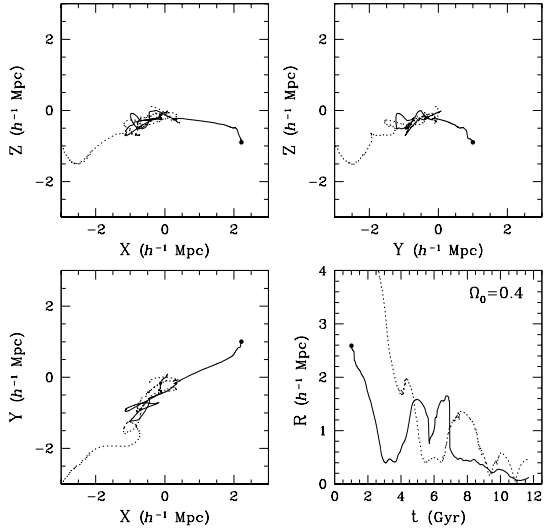}
\caption{Three projections of galaxy trajectories in the
  $\Omega_0=0.4$ cluster.  Notation as in Figure
  \protect\ref{fig:traject_om1}.  Solid line is for the galaxy ranked
  \#4, dotted line is for the galaxy ranked \#93.
  \label{fig:traject_om.4}}
\end{figure}

\begin{figure}
\plotone{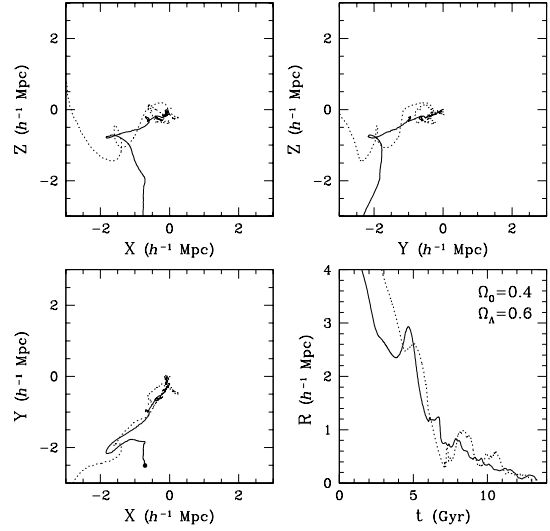}
\caption{Three projections of galaxy trajectories in the
  $\Omega_0=0.4$, $\Omega_\Lambda=0.6$ cluster.  Notation as in Figure
  \protect\ref{fig:traject_om1}.  Solid line is for the galaxy ranked
  \#4, dotted line is for the galaxy ranked \#87.
  \label{fig:traject_om.4L}}
\end{figure}

Figures \ref{fig:traject_om1}--\ref{fig:traject_om.4L} show the
trajectories of two selected halos in each cluster.  These galaxies
are used for the high resolution simulations in Paper II, representing
a large spiral and a dwarf spheroidal, respectively.  Their
trajectories are quite irregular.  In the $\Omega_0=1$ case, the
galaxies are just entering the cluster at the present epoch, although
not on a free-fall path.  The orbits are similar to those shown in
\cite{Fetal:96}.  On the other hand, in the low $\Omega_0$ cases the
galaxies have enough time to move around the cluster center several
times, more so in the open model.  These differences will have major
effects on the dynamical evolution of the galaxies.

In general, there is a good agreement between the trajectories
calculated using two different methods, especially for the less
massive galaxies.  For the very massive halos, the ``test particle''
motion occasionally tends to oscillate around the trajectory
determined by the densest particle subset.  I have therefore decided
to use the latter as the more reliable trajectories.

\section{Mergers and Close Encounters}
  \label{sec:gal_distr}

The hierarchical formation of clusters and the survival of
substructure lead to more frequent collisions of galaxies and lower
relative velocities at early epochs.  Preferentially radial infall
implies that the final distribution of galaxies is very dense.  Most
of the galaxies lie close to the cluster center, while the rest form a
scarcely populated halo.

The number density of massive galaxies ($M_{\rm min} \approx 10^{11}\,
M_{\sun}$) can be fitted by a power law with a central core:
\begin{equation}
n_g(r) = {n_0 \over (r + r_0)^\gamma},
  \label{eq:gal_den}
\end{equation}
normalized to the total number of galaxies $N_g$ within the virial
radius of the cluster, $R_{\rm vir}$.  Not all initially identified
halos end up in the clusters, so with $N_g < 100$ trying to estimate
the core radius $r_0$ and the slope $\gamma$ by fitting a histogram of
galaxy positions yields highly uncertain results.  Instead, I use a
maximum likelihood (ML) method analogous to that in \cite{G:97}.  The
parameters $r_0$ and $\gamma$ are to be determined by searching for
the maximum of the ML function, ${\cal L} = e^{-N_g} \prod_{i=1}^{N_g}
4\pi r_i^2 n(r_i)$.  Requiring the two partial derivatives of
$\log{\cal L}(r_0,\gamma)$ to vanish gives a system of two integral
equations, which have been solved numerically using the software
package Maple.  The resulting parameters are given in Table
\ref{tab:gal}.

The slopes of the density profile are steep, $\gamma = 2.4 - 4.2$.
The massive galaxies are more concentrated in the inner parts of the
clusters than the dark matter which follows an approximately
isothermal profile, $\rho_{dm} \propto r^{-2}$.  Also, the core radii
in all three clusters are very small, below the force resolution
scale.  At such small distances from the center the distribution of
galaxies can be affected by the giant cD galaxy which may swallow its
neighbors.  To check this, I have repeated the ML analysis excluding
all galaxies within one resolution element, $\Delta$, from the center.
This forces the core radii to be at least one $\Delta$.  The resulting
values (Table \ref{tab:gal}) are very close to this limit, suggesting
that the true core radii can indeed be quite small.  Thus the effect
of the central giant on the overall galaxy distribution seems to be
small.

\subsection{Number of Close Encounters}

The detailed trajectories of galaxies can be used to calculate the
number of close encounters during the epoch of cluster formation.  In
order to detect all encounters, the trajectories are linearly
interpolated between the successive time steps.  The comoving
separations are converted into physical kiloparsecs using the
appropriate value of the Hubble constant at each time step.  If a
galaxy is found to have merged (see Appendix \ref{sec:app_merger}),
its subsequent trajectory is ignored.

Table \ref{tab:enc} shows the number of encounters, $N_{\rm enc}$, as
a function of the impact parameter, $R_{\rm enc}$.  It is defined as a
minimum distance between the centers of mass of the galaxies.  The
last column gives the ``total'' number of encounters for $R_{\rm
enc}=40$ kpc (it is assumed that no mergers are possible for the
separations larger than 40 kpc).  Table \ref{tab:enc} also shows the
average number of close encounters per galaxy, $N_{\rm enc,1}$.  Since
each encounter involves two galaxies, this number is $N_{\rm enc,1}
\equiv 2\, N_{\rm enc}/N_g$.

Figure \ref{fig:enctime} shows the number of collisions per galaxy
within $R_{\rm enc} = 20$ kpc.  Note that in the low $\Omega_0$
clusters, most collisions take place near the end of the simulation,
when the galaxies have assembled inside the cluster.  In Cluster I,
the galaxies are closer to each other initially and have close
encounters uniformly distributed in time.  Several distinct peaks,
notably in Clusters I and III, can be due to the new groups of
galaxies entering the clusters.

The number of close encounters does not scale very well with $R_{\rm
enc}^2$, as predicted by equation (\ref{eq:enc}).  Since most
encounters take place in the already formed cluster, any mergers prior
to that remove prospective galaxies and reduce the number of
subsequent collisions.  To test that equation (\ref{eq:enc}) would
have been valid in the absence of mergers, I count the number of
encounters disallowing mergers.  The new numbers, $\tilde{N}_{\rm
enc}$, scale as $R_{\rm enc}^2$ in agreement with the prediction (see
Table \ref{tab:enc}).

However, the total number of encounters is much higher than expected.
The effect is mainly due to the concentration of galaxies at the
center of the cluster, implying a smaller effective volume.
Equation (\ref{eq:enc}) can be rewritten in a general form:
\begin{equation}
N_{\rm enc} = \int \pi R_g^2 \, V_{\rm enc} \, t_H \, n_g^2(r) \, dV,
\end{equation}
where the integration extends over the virial volume of the cluster, $V$.
The mean value of the number density squared, $\langle n_g^2 \rangle$,
is substantially higher than the square of the mean density, $\langle
n_g \rangle^2$, because of the galaxy clustering.  Their ratio is the
clumping factor
\begin{equation}
{\cal C} \equiv {\langle n_g^2 \rangle \over \langle n_g \rangle^2}
  = {V \, \int n_g^2(r) \, dV \over \left( \int n_g(r) \, dV \right)^2}.
\end{equation}
Using the azimuthally-averaged density profile (\ref{eq:gal_den}) and
the auxiliary functions
\begin{equation}
f_n(r_0,\gamma) \equiv
  \int_0^1 {x^2 \, dx \over (x + r_0/R_{\rm vir})^{n \gamma}},
\end{equation}
I find ${\cal C} = {f_0 f_2 / f_1^2}$ (see Table \ref{tab:gal}).
Now, the number of close encounters is
\begin{equation}
N_{\rm enc} = 320 \, \left( {{\cal C} \over 1000} \right) \,
          \left( {N_g \over 65} \right)^2 \,
          \left( {R_{cl} \over 2\ {\rm Mpc}} \right)^{-3} \,
          \left( {R_{\rm enc} \over 10\ {\rm kpc}} \right)^2 \,
          \left( {V_{\rm enc} \over 800\ {\rm km\ s}^{-1}} \right).
  \label{eq:enc2}
\end{equation}
This is much higher than the naive estimate (\ref{eq:enc}) and agrees
well with the results for Cluster III.  

Because of the higher degree of clustering in the low $\Omega_0$ models,
more than half of all encounters occur within the inner 0.4 Mpc from
the center.  In contrast, in the $\Omega_0=1$ model collisions take
place throughout the cluster, leading to a smaller value of the
clumping factor.

\subsection{Relative Velocities}
  \label{sec:relvel}

Figure \ref{fig:encvel} shows the relative velocities of galaxies at
the time of collision.  The distributions are skewed, peaking at low
values and extending large tails into the high velocities.  A
characteristic value, the median, is around 800 km s$^{-1}$.  Again,
there is a strong difference between Cluster I and the low $\Omega_0$
clusters.  Table \ref{tab:relvel} demonstrates that while the peak
velocity is higher in Cluster I, the median and mean values are
substantially lower than those in Clusters II and III.  The very high
velocities are almost unattainable in the $\Omega_0=1$ model.

Table \ref{tab:disp} shows that the velocity dispersion of galaxies is
smaller than that of dark matter and varies strongly with the position
in the cluster.  Due to a small number of identified galaxies I have
split them in three equally-populated radial bins.  In Clusters II and
III, the velocity gradient is strong and inverted, with the lowest
dispersion within the inner cluster core.  In Cluster I the effect is
less conspicuous.  Since the samples are small, Table \ref{tab:disp}
serves only to indicate the trend.  In contrast, ``test particle''
galaxies tracing the dark matter potential show a normal gradient
weakly rising towards the central bin (from 550 km s$^{-1}$ to 750 km
s$^{-1}$ in Cluster II).  Thus the inverted gradient is a property of
only real galaxies.

A review of the literature reveals that numerical simulations often
produce ``galaxies'' with the pairwise velocity dispersion lower than
that of dark matter (\citealt{CKK:00,SDE:95,C:94}; but see also
\citealt{OH:99,Detal:99,GHetal:00}).  In a specific study of the
relative velocities at the time of galaxy collisions, \cite{TDS:98}
find that the peak velocity is only $V_{\rm rel} \approx 2.5 \,
\sigma_g \approx 500$ km s$^{-1}$ (their Fig. 7).  There are several
reason for this: (i) A statistical bias of hierarchical structure
formation, whereas galaxies form around highly-clustered density
peaks.  Therefore, large galaxies are always around the cluster core
and do not gain velocity from gravitational infall.  (ii) A physical
bias relating to gas cooling and star formation \citep{CO:00} makes
younger galaxies move more slowly than the older ones.  (iii)
Dynamical friction reduces the angular momentum of large galaxies and
draws them towards the center.  In the present simulations, the final
distances from the center are anti-correlated with galaxy masses
(Spearman's correlation coefficient is $r_s \approx -0.3$ for Clusters
II and III, but only $r_s \approx -0.1$ for Cluster I).  Thus there is
a moderate amount of mass segregation in the clusters formed at
earlier times.

Finally, numerical effects may lead to low galactic velocities.  The
PM code loses force resolution in the two inner grid cells, within
which the galaxies do not ``feel'' each other's gravity.  Also, if the
halo particles departed from its center, the center-of-mass velocity
would be significantly reduced.  However, if they were scattered
completely randomly in the cluster the average velocities would have
fallen below 100 km s$^{-1}$, which is clearly ruled out.

At present, observations neither confirm nor reject the velocity bias.
\cite{CYE:97}, in a survey of 1150 galaxies from 14 clusters, find an
almost constant projected velocity dispersion up to the virial radius.
The line-of-sight projection may, however, wash out a decrease of the
true velocity dispersion near the center.  Another large survey of 15
rich clusters \citep{DS:88} shows a significant amount of substructure
with the lower pairwise velocities of galaxies, as well as a modest
decline (of the order 20\%) of the azimuthally averaged dispersion in
the central regions.  So the current observational status of the
velocity bias is inconclusive.

\subsection{Merger History}
  \label{sec:merger}

The low relative velocities of galaxies make mergers more likely.
Without having enough resolution to study mergers in detail, I adopt a
simple prescription based on the impact characteristics of the
galaxies.  It consists of the three merger criteria described in
Appendix \ref{sec:app_merger}.

Figure \ref{fig:enctime} and Table \ref{tab:enc} show the expected
number of mergers per galaxy, $N_{\rm mer,1}$.  It is calculated as
$N_{\rm mer}/N_g$, since each merger removes only one galaxy.  The
calculated merger probability is high.  At impact parameters $R_{\rm
enc} < 10$ kpc it is already $N_{\rm mer,1} \sim 10\%$, and the total
value reaches $\gtrsim 20\%$.  Most of the mergers take place shortly
after galaxies enter the clusters (even earlier in Cluster I).  Once
the clusters virialize, galactic velocities become too high and
prohibit mergers.  Yet, the number of mergers is only a very small
fraction, about 0.1\%, of the number of close encounters (cf
eq. [\ref{eq:mer}]).

\cite{GHetal:98} find very few mergers within the virial radius of
their high-resolution cluster.  This can be reconciled with our
results because \cite{GHetal:98} count only the recent encounters in
the virialized cluster ($z < 0.5$) while our numbers include all
encounters starting from $z_g=5$.  Note that most mergers occur at
relatively early times (Figure \ref{fig:enctime}).  In the three cases
considered, the last merger occurred at 2.3, 1.0, and 3.3 Gyr before
the present, respectively.  The merger probability of 22--29\% may
reflect the instances of the formation of large elliptical galaxies at
high redshifts.  They are expected to take place naturally in the
centers of future clusters.

\section{Tidal Field Around Galaxies}
  \label{sec:tidal}

The tidal field of the cluster is calculated along the trajectories of
the identified galaxies.  The tidal force $F_{\rm tid,\alpha}$ and the shear
tensor $F_{\alpha\beta}$ are
\begin{equation}
F_{\rm tid,\alpha} \equiv 
  - \left( {d^2\Phi \over dr_\alpha dr_\beta} \right)_0 \, r_\beta
  \equiv F_{\alpha\beta} \, r_\beta,
\end{equation}
where $\Phi$ is the external potential and $\rr$ is the radius-vector
in the galactic reference frame.  By definition tidal force is zero at
the center of mass of the galaxy.  I evaluate the tidal tensor at each
time step for each identified halo, using the tri-linear interpolation
scheme with a smoothing filter described in Appendix
\ref{sec:app_tidal}.

Being coordinate-independent, the trace of the tidal tensor serves as
a good measure of the tidal field.  Figures
\ref{fig:tidal_om1}--\ref{fig:tidal_om.4L} show the trace
$F_{\alpha\alpha}$ around the pairs of large and dwarf galaxies chosen
in \S\ref{sec:traject} from each cluster simulation.  The time
variation of the tidal force is extremely irregular.  There are many
strong peaks, reaching the amplitude of 200 Gyr$^{-2}$ and higher.
The largest peaks are usually narrow, with the characteristic duration
of $10^8$ yr, while the smaller peaks last longer.  The occurrence of
the two types of peaks is clearly differentiated in time: the larger
happen early, when massive galaxies are distinctly denser than the
background; the smaller happen at later stages, when the halos merge
and the gravitational potential becomes smoother.

The amplitude of the tidal field is the largest in the $\Omega_0=0.4$
simulation, although strong perturbations continue for a longer time
in the $\Lambda$ model.  In the $\Omega_0=1$ model, the overall
amplitude is smaller but there is more temporal variation.  The main
distinction between the galaxies in the low and high $\Omega_0$ models
is the time spent within the cluster.  In the latter case, the
galaxies are just arriving to the center, while in the low $\Omega_0$
models the galaxies have entered earlier and experienced more tidal
interaction.

The overall amplitude of the tidal force is determined by the epoch of
cluster formation.  For the same virial overdensity, the cluster is
denser on average when the mean density of the Universe is high.  In
the low $\Omega_0$ models clusters form at a higher redshift and
therefore exert stronger tidal forces than in the $\Omega_0=1$ model.
Since the subsequent growth of the clusters is balanced by the drop of
the mean density, the tidal force is expected to be of the same order
of magnitude throughout the evolution.

Note that the peaks of the tidal force are not always correlated with
closest approach to the cluster center.  The galaxies are being
tidally ``shocked'' at 1 Mpc from the center as often as at 400 kpc.
Even though the center of the cluster has been moving slowly in the
simulation box, it is still likely that close encounters with massive
galaxies or groups of galaxies happen relatively far from the center.
As the cluster grows, such tidal encounters occur within a fairly
large (comoving) volume.

The amplitude of the tidal force tells us about the effective scale of
perturbation.  For a smooth cluster, the tidal tensor is of the order
\begin{equation}
F_{\alpha\beta} \sim {G M(r)\over r^3} \sim {\sigma_{cl}^2 \over r^2},
\end{equation}
where $M(r)$ is the enclosed mass within radius $r$.  In the simulated
clusters the three-dimensional velocity dispersion is $\sigma_{cl} =
\sqrt{3} \times 660$ km s$^{-1}$.  If a characteristic scale, the
cluster core radius, is 200 kpc then the expected tidal perturbation
is
\begin{equation}
F_{\alpha\beta} \sim 32 \; \left({\sigma_{cl} \over 1100\
  {\rm km\ s}^{-1}}\right)^2
  \, \left({r \over 200\ {\rm kpc}}\right)^{-2} \; {\rm Gyr}^{-2}.
\end{equation}
The measured amplitude of the tidal force (cf. Figures
\ref{fig:tidal_om1}--\ref{fig:tidal_om.4L}) is about 150 Gyr$^{-2}$.
Turning the argument around, such tidal force corresponds to the
effective scale of perturbation
\begin{equation}
r_p = 92 \; \left({\sigma_{cl} \over 1100\ {\rm km\ s}^{-1}}\right) \,
      \left({F_{\alpha\beta} \over 150\ {\rm Gyr}^{-2}}\right)^{-1/2} \;
      {\rm kpc}.
\end{equation}
Smaller than the core radius but larger than individual galaxies, this
scale is appropriate for small groups of galaxies.  As discussed in
\S\ref{sec:obs}, groups maintain their identity in many clusters.
Even such a relaxed cluster as Coma shows a detectable amount of
substructure \citep{CD:96,CG:99}.  Since large peaks of the tidal
force dominate the total amount of heating (see \S\ref{sec:heating}),
I conclude that the tidal effects are produced primarily by the
interaction of galaxies with the remaining substructure in clusters.

\section{Tidal Heating}
  \label{sec:heating}

The main effect of external tidal forces on a galaxy is a positive
energy change of stars and dark matter, or {\it tidal heating}.  The
amount of tidal heating can be estimated analytically, at least for
the spherically-symmetric component.  For a single tidal shock, the
energy input is the ensemble average of the velocity change squared,
${1\over 2} \langle (\Delta v)^2 \rangle$, where $\Delta v$ is a time
integral of the tidal force times the reduction factor that takes into
account the conservation of adiabatic invariants of stellar orbits
\citep{GHO:99}.  In the simulations, the tidal force has many peaks of
variable duration.  In addition, different components of the tidal
tensor are not perfectly correlated so that they experience maxima and
minima at different times.  Clearly, this complex behavior cannot be
fully described as simply as a single tidal perturbation.  Instead, I
design a cumulative parameter, based on the semi-analytic theory of
tidal shocks, which provides a useful scaling relation for the amount
of tidal heating.

Assume that each peak of the tidal force can be considered a separate
``tidal shock''.  The total amount of heating is then the sum of the
contributions from all peaks.  Moreover, since the velocity changes
add in quadrature, all components of the tidal tensor can be treated
separately.  Then it follows from the semi-analytic theory
(cf. \citealt{GO:99}) that the total energy per unit mass of the
system changes by
\begin{equation}
\langle\Delta E\rangle = {1\over 6} \, I_{\rm tid} \, 
   \langle r^2 \rangle,
  \label{eq:de_tid}
\end{equation}
where $r$ is a characteristic size of the galaxy and $I_{\rm tid}$ is
the tidal heating parameter
\begin{equation}
I_{\rm tid} \equiv \sum_n \sum_{\alpha,\beta} \left( \int
  F_{\alpha\beta} \, dt \right)_n^2 \, 
  \left( 1 + {\tau_n^2 \over t_{\rm dyn}^2} \right)^{-3/2},
  \label{eq:ipar}
\end{equation}
where the sum extends over all peaks $n$ and all components of the
tidal tensor, $\alpha,\beta=\{x,y,z\}$.  Here $\tau_n$ is the
effective duration of peak $n$ for each value of $\alpha$ and $\beta$,
and $t_{\rm dyn}$ is the half-mass dynamical time.  For the galaxy
model from Appendix \ref{sec:app_merger}, the dynamical time scales
with the halo mass as $t_{\rm dyn} \propto M_g^{1/4}$.  The inner
integration is limited to the time intervals between the successive
minima of the tidal force.  In carrying out the integration, I have
linearly interpolated $F_{\alpha\beta}$ between the output points to
match the time step of the galactic simulations presented in Paper II.
This provides the most direct way to compare the analytical estimate
$I_{\rm tid}$ with the results of self-consistent \N-body simulations.
The agreement is generally good.

Figure \ref{fig:tidalsum} shows the distribution of parameters $I_{\rm
tid}$ for the three clusters.  There are more galaxies with the low
value of $I_{\rm tid}$ than with the high -- tidal effects are
different for different galaxies.  The histograms display a distinct
trend: the low $\Omega_0$ clusters have stronger tidal heating than
the $\Omega_0=1$ cluster.  A least-squares fit to the histograms,
using a convenient functional form $N(I_{\rm tid}) = N_0 \,
\exp(-I_{\rm tid}/I_0)$, yields the following characteristic
parameters:
\begin{equation}
\begin{array}{ll}
I_0 = 8 \times 10^2 \ \mbox{Gyr}^{-2} & \;\;\;\; \mbox{(Cluster I),} \\
I_0 = 5 \times 10^3 \ \mbox{Gyr}^{-2} & \;\;\;\; \mbox{(Cluster II),} \\
I_0 = 2 \times 10^3 \ \mbox{Gyr}^{-2} & \;\;\;\; \mbox{(Cluster III).}
\end{array}
  \label{eq:I0}
\end{equation}
The timescale of perturbation, $\tau_n$, is very important in
determining the total amount of heating.  Without the adiabatic
correction factor, the values of $I_0$ would have been at least an
order of magnitude higher.  Overall, the open $\Omega_0=0.4$ model has
the highest heating rates and the $\Omega_0=1$ model the lowest.
Thus galaxies should experience considerably more dynamical evolution
in a low density universe than in the $\Omega_0=1$ case.

A quick estimate whether tidal heating can turn a disk galaxy into an
elliptical is the increase of the velocity dispersion induced by the
energy change (eq. [\ref{eq:de_tid}]): ${1\over 2} \sigma_{\rm tid}^2
\equiv \langle\Delta E\rangle$.  For a typical tidal parameter $I_{\rm
tid} = 10^3$ Gyr$^{-2}$, the inner region at 3 kpc will be heated to
$\sigma_{\rm tid} = 54$ km s$^{-1}$.  This is not enough to completely
randomize stellar velocities in a large galaxy, but enough to turn a
thin disk into a thick one (cf. \S\ref{sec:nbody}).  In order to reach
the velocity dispersion of a large elliptical galaxy, 250 km s$^{-1}$,
the tidal heating parameter would need to be $2.2\times 10^4$
Gyr$^{-2}$.  This is above the maximum for our clusters, and therefore
the ellipticals are unlikely to be created by tidal interactions.

\subsection{Tidal Effects from Close Encounters}

The smoothing filter used in the calculation of the tidal field
(Appendix \ref{sec:app_tidal}) automatically removes the effect of
close galactic encounters.  Their additional contribution to the tidal
heating rate can be estimated as follows.  The number of close
encounters per galaxy in the simulation as a function of impact
parameter $R = R_{\rm enc}$ is (cf. eq. [\ref{eq:enc2}])
\begin{equation}
N_{\rm enc,1} \approx 10\, \left({R \over 10\, \mbox{kpc}}\right)^2 \,
                      \left({V_{\rm enc} \over 800\ {\rm km\ s}^{-1}} \right).
\end{equation}
The amplitude of the tidal force due to a galaxy of mass $M_g$ is of
the order $F_{\alpha\beta} \sim G M_g(R)/R^3$.  The duration of the
encounter is approximately $R/V_{\rm enc}$, so that the contribution
to the heating parameter at a distance $R$ is
\begin{equation}
I_{\rm enc}(R) = \left( \int \, F_{\alpha\beta} \, dt \right)^2
           \approx \left({G \, M_g(R) \over R^2 \, V_{\rm enc}}\right)^2.
\end{equation}

The total contribution of all encounters is the integral over all
impact parameters weighted by the number of encounters, $N_{\rm
enc,1}$.  The perturbing halo can be represented by a truncated
isothermal sphere with the density profile given by equation
(\ref{eq:app_den}).  The lower limit of integration is set by the
condition $N_{\rm enc,1} > 1$, i.e. that at least one encounter takes
place; this gives $R_{\rm min} \approx 3$ kpc.  The upper limit can be
taken as the tidal radius, $R_{\rm max} \sim 30$ kpc, but in the end
the heating rate depends only logarithmically on the integration
limits.
\begin{eqnarray}
I_{\rm enc} & = & \int_{R_{\rm min}}^{R_{\rm max}} I_{\rm enc}(R) \,
                  dN_{\rm enc,1}(R) \nonumber\\
        & \approx &
          4.9 \times 10^2 \,
          \left({\sigma_g \over 200 \, \mbox{km s}^{-1}} \right)^4 \,
          \left({V_{\rm enc} \over 800 \, \mbox{km s}^{-1}} \right)^{-1} \
          \mbox{Gyr}^{-2},
  \label{eq:ienc}
\end{eqnarray}
where $\sigma_g$ is the galactic velocity dispersion.  This value of
$I_{\rm enc}$ is comparable to, but smaller than, the characteristic
parameter $I_0$.  Thus close encounters contribute an important part
but do not dominate the tidal heating.

\subsection{Self-consistent \N-body simulations}
  \label{sec:nbody}

These analytical estimates will be substantiated in Paper II
\citep{PaperII} which presents self-consistent \N-body simulations of
several galaxies selected from each cluster sample.  Each galaxy is
resimulated with $2\times 10^6$ particles, supplementing the internal
dynamics with the external tidal forces of the cluster.  Disks of
large spiral galaxies survive tidal heating but thicken in the
vertical direction by a factor of two, while their dark matter halos
are truncated at roughly 30 kpc for $\sigma_g = 250$ km s$^{-1}$.  At
the end of the simulation they resemble featureless S0 galaxies.  In
contrast, low surface brightness galaxies are almost completely
disrupted by tidal forces.

Most of the results can be explained qualitatively by the critical
tidal density, corresponding to the trace of the tidal tensor via
Poisson's equation: $4\pi G \rho_{\rm tid} \equiv F_{\alpha\alpha}$.
It is similar to the instantaneous density of the Roche lobe, but
reflects the highest tidal peaks along the galactic orbit.  This
density $\rho_{\rm tid}$ determines the truncation radius of the halos
and the amount of luminous and dark mass stripped by tidal heating.

Galactic \N-body simulations confirm that the amount of tidal heating
is set by $I_{\rm tid}$ (or $\rho_{\rm tid}$).  The evolution of
galaxies is stronger on average in Cluster II.  However, reflecting
the wide distribution of tidal parameters in each cluster (Fig.
\ref{fig:tidalsum}), the heating on different orbits within the same
cluster varies as much as between the different cluster models.

\section{Conclusions}

Tidal interactions may explain the observed dynamical and
morphological evolution of galaxies in clusters.  Using the
simulations of hierarchical cluster formation, I calculate the tidal
forces along the galactic trajectories starting at redshift $z_g = 5$.
The tidal field shows a lot of temporal variation.  The peaks of the
tidal force do not always correlate with the closest distance to the
cluster center and are likely to be due to the infalling groups of
galaxies or individual massive galaxies.  The tidal peaks are stronger
and more impulsive in the low $\Omega_0$ models.  Also, in those
models the cluster forms at earlier epochs and the galaxies experience
more interaction.  Thus overall tidal effects are stronger in the low
$\Omega_0$ clusters.

Massive galaxies are strongly clustered at the end of the simulations.
Together with the remaining substructure, this significantly increases
the probability of galactic collisions.  The number of encounters
within 10 kpc is about 10 per galaxy in the Hubble time.  About 0.1\%
of such encounters is expected to result in a merger, defined by the
analytical merger criteria in \S A.  This gives a high merger
probability of the identified halos, 22--29\% per halo, although most
mergers take place before the clusters virialize.

The amount of dynamical evolution of galaxies induced by tidal
interactions can be estimated using the tidal heating parameter,
$I_{\rm tid}$.  The typical values of $I_{\rm tid} \sim 10^3$
Gyr$^{-2}$ result in an increase of the stellar velocity dispersion by
$\sim 50$ km s$^{-1}$.  At present this is insufficient to turn a
large disk galaxy into an elliptical galaxy.  However, in the future a
further secular evolution and transformation into the ellipticals
seems inevitable.  Close galactic encounters cannot be resolved in
present simulations, but they are estimated to contribute a fraction
of 10\% to 50\% of the total heating rate.

\acknowledgements

I would like to thank my thesis advisor J. P. Ostriker for his support
and guidance, Nick Gnedin for expertise in cosmology, Renuye Cen for
his PM code, Greg Bryan for sharing his AMR code, Mike Blanton for his
excellent Points visualization software, and Jeremy Goodman, David
Spergel, Neta Bahcall, and Scott Tremaine for valuable comments.  I
acknowledge the support from NSF grants AST 93-18185 and AST 94-24416.
This work was submitted in partial fulfillment of the
Ph.D. requirements at Princeton University.


\appendix
\section{Merger Criteria and Merging Time}
  \label{sec:app_merger}

This appendix describes the merger criteria used in \S
\ref{sec:merger}.  The physical processes leading to merger are the
gravitational capture of two galaxies and the subsequent dynamical
friction of their halos.  The energy exchange during the first
encounter determines whether the galaxies form a bound pair, while
dynamical friction determines the timescale for final coalescence.
Also, the tidal forces of the cluster can disrupt even a bound system
at its point of maximum separation.

For clarity, I model the halos as truncated isothermal spheres with
the parameters scaled to the initial galaxy mass, $M_i$.  Assume that
the velocity dispersion scales as $\sigma_g = \sigma_*
(M_i/M_*)^{1/4}$, in agreement with the Tully-Fisher and Faber-Jackson
relations.  Here $\sigma_* = 250$ km s$^{-1}$ and $M_* = 2\times
10^{12}\, M_{\sun}$ are the fiducial circular velocity and mass of an
$L_*$ galaxy.  Furthermore, assume that the velocity dispersion does
not change in the course of evolution but the density distribution is
truncated at a radius $R_t = R_{t,*} \, \sigma_g/\sigma_*$.  The
latter assumption is confirmed by the self-consistent \N-body
simulations presented in Paper II, and $R_{t,*}$ is determined from
the simulations.  The density profile of the halos is
\begin{equation}
\rho(r) = {\sigma_g^2 \over 4\pi G (r^2 + R_c^2)},
   \hspace{1cm} \mbox{for}\ r < R_t
   \label{eq:app_den}
\end{equation}
where $R_c$ is the core radius.  The enclosed mass scales almost
linearly with $R_t$.  Let galaxy 1 be more massive initially than
galaxy 2; it will still be more massive after the truncation, $M_{1}
\geq M_{2}$.  As usual, the two-body problem can be written in terms
of the reduced particle with the mass $\mu \equiv M_1 M_2 /(M_1+M_2)$.
The dynamics is characterized by the radius of encounter, $\RR_{\rm
enc} = \rr_1 - \rr_2$, and the relative velocity, $\VV_{\rm enc} =
\vv_1 - \vv_2$.

\subsection{Orbital Energy}

Dynamical friction works when the halos overlap.  For that the
distance of closest approach during the first encounter, $R_{\rm
enc}$, must be smaller than the tidal radius of the larger halo,
$R_{t,1}$.  Then the potential energy of the binary system is given by
\begin{equation}
W = \int \Phi_1(\rr + \RR_{\rm enc}) \rho_2(\rr) d\rr
  = -{\sigma_1^2 \, M_2 \over 1-\chi_2} \, {\cal I},
  \label{eq:W}
\end{equation}
where 
$\chi_2 \equiv \gamma_2 \arctan{\gamma_2^{-1}}$,
$\gamma_2 \equiv R_{c,2}/R_{t,2}$ and
\begin{equation}
{\cal I} = {1 \over 2} \int_0^1 {x^2 dx \over x^2 + \gamma_2^2}
   \int_0^\pi \sin{\theta} d\theta \left[ 1 - {R_{c,1} \over 
   |\rr+\RR_{\rm enc}|}
   \arctan{|\rr+\RR_{\rm enc}| \over R_{c,1}} + {1\over 2}
   \ln{R_{t,1}^2 + R_{c,1}^2 \over (\rr+\RR_{\rm enc})^2 + R_{c,1}^2} \right],
\end{equation}
where $x = r/R_{t,2}$.  The integral ${\cal I}$ varies slowly with
$R_{\rm enc}$ and is always of order unity.  In the limit $R_{c,1},
R_{t,2} \ll R_{t,1}$ and also in the case $R_{t,1} \approx R_{t,2}$,
the integral evaluates to ${\cal I} \approx (1-\chi_1) (1-\chi_2)$.
Thus the orbital energy of the reduced particle is
\begin{equation}
E_{\rm orb} \approx {1\over 2} \mu \, V_{\rm enc}^2 - M_2 \,
                     \sigma_1^2 (1-\chi_1).
\end{equation}
Note that it does not depend on the radius of encounter as long as the
two halos overlap.  Also, being enclosed within the larger halo, the
smaller halo is further truncated according to $R_{t,2} \approx R_{\rm
enc} \, \sigma_2/\sigma_1$.

The dissipation of orbital energy into the internal stellar motion can
produce a bound orbit even if initially $E_{\rm orb} > 0$.  This is a
non-linear interaction and cannot be calculated analytically.
Instead, an approximate functional dependence can be derived and then
fitted to the results of \N-body experiments.  The velocity drag
during the encounter follows from the dynamical friction analysis
(\citealt{BT:87}; also S. Tremaine, 1998 private communication):
\begin{equation}
{dV_{\rm enc} \over dt} \sim {4\pi G^2 M_2 \rho_1 \over V_{\rm enc}^2},
  \label{eq:df}
\end{equation}
where the slow-varying Coulomb logarithm is neglected.  The encounter
lasts of the order $\Delta t \sim R_{\rm enc}/V_{\rm enc}$, so that
the energy change is
\begin{equation}
\Delta E \sim \mu V_{\rm rel} \, {dV_{\rm enc} \over dt} \, \Delta t
         \sim \mu \, {M_2 \over M_1} \,
              {\sigma_1^4 \over V_{\rm enc}^2 + \sigma_1^2}.
  \label{eq:de_orb}
\end{equation}
The last factor assures that equation (\ref{eq:de_orb}) remains valid
even in the limit where the relative velocity is lower than the
velocity dispersion of the larger galaxy, $V_{\rm enc} \lesssim
\sigma_1$.

The normalization is provided by the simulations of \cite{MH:97} who
considered a hyperbolic encounter of two equal-mass galaxies.  I
derive $\Delta E$ as the initial orbital energy leading to a bound
pair of galaxies after the first passage.  For the encounters with the
impact parameter close to the galactic tidal radius, the correction
factor for equation (\ref{eq:de_orb}) is about 2.4.  Thus the galaxies
become bound if they satisfy the {\it capture criterion}:
\begin{equation}
E_{\rm orb} < \Delta E = 2.4 \, \mu \, {M_2 \over M_1} \,
                     {\sigma_1^4 \over V_{\rm enc}^2 + \sigma_1^2}.
  \label{eq:mercond}
\end{equation}

\subsection{Dynamical Friction Time}

Even if two galaxies satisfy the capture criterion, they may not have
enough time to merge until the present.  To estimate the merger
timescale via dynamical friction, assume that the velocity of the
reduced particle is $v \approx (\sigma_1^2 + \sigma_2^2)^{1/2}$.  For
circular orbits of radius $r$, the rate of loss of the specific angular
momentum is given by the Chandrasekhar formula (eq. [\ref{eq:df}]):
\begin{equation}
{dJ \over dt} \approx r {dv \over dt} \sim
   - {M_2 \over M_1} \, {\sigma_1^4 \over v^2}.
  \label{eq:dfj}
\end{equation}
For eccentric orbits, $r$ should be taken as the distance of closest
approach.  Recent work extends Chandrasekhar's calculation and
includes a time-dependent response to the perturbation
\citep{SD:94,CP:98,DTGF:98}.  However, the complicated dynamics of the
interaction cannot be easily parametrized.  Instead, I employ the
simple functional form given by equation (\ref{eq:dfj}) and use
numerical simulations again to obtain correct normalization.  Taking
now $dJ/dt \approx v \, dr/dt$, yields the inspiraling time
\begin{equation}
t_{DF} \sim {r \over \sigma_1} \, \left( {v \over \sigma_2} \right)^3.
  \label{eq:dftime}
\end{equation}
The normalization follows from the simulations of \cite{DMH:99} who
studied the formation of tidal tails by the interacting equal-mass
galaxies.  From the merging time of their models (their Table 3), I
deduce $t_{DF} \propto r$ and $t_{DF} \propto \sigma_1^{-0.5}$,
roughly in agreement with equation (\ref{eq:dftime}).  The resulting
timescale is
\begin{equation}
t_{DF} = 0.6 \left({r \over 10\ \mbox{kpc}}\right) 
             \left({\sigma_1 \over 250\ \mbox{km s}^{-1}}\right)^{-1}
             {1 \over \eta^3} \left({1+\eta^2 \over 2}\right)^{3/2} \,
         \mbox{Gyr},
  \label{eq:mercond2}
\end{equation}
where $\eta \equiv \sigma_2/\sigma_1$.  If the time of collision
$t_{\rm coll}$ does not satisfy the {\it dynamical friction criterion}
$t_{\rm coll} + t_{DF} < t_H$, the merger is disallowed.

\subsection{Tidal Disruption}

According to numerical simulations of isolated mergers \citep{M:99},
after a first encounter the galaxies separate on a highly elongated
orbit before coming back for a final unification.  In the cluster, the
external tidal force at the distance of maximum separation can
prevail the gravitational attraction of the two galaxies and disrupt
the pair.

Tidal disruption in clusters is important for the following reason.
Halos are truncated at the radius $R_{t,1}$ where the tidal force
balances the gravitational pull of the galaxy.  If after the encounter
the smaller galaxy moves away more than $R_{t,1}$ from the center of
the larger galaxy, the external tidal field would be stronger than the
mutual attraction.

There is still a possibility, however, that the pair will merge.  The
tidal field varies along the galaxy trajectory and the truncation is
determined by the peak values.  If the collision takes place when the
external tidal field is weak, the galaxies may remain bound.

Quantitatively, the distance of maximum separation $r_m$ can be
determined from the orbital energy, which must now be negative after
including the dissipated term (eq. [\ref{eq:mercond}]).  If the halos
detach completely after the encounter,
\begin{equation}
E_{\rm orb} = {\mu \, J^2 \over 2 r_m^2} - {G M_1 M_2 \over r_m} < 0.
  \label{eq:Eorbsep}
\end{equation}
The angular momentum term can usually be ignored, so that the maximum
distance is
\begin{equation}
{r_m \over R_{t,1}} \approx (1-\chi_1) \, (1+M_2/M_1) \,
                      {\mu \sigma_1^2 \over |E_{\rm orb}|}.
  \label{eq:rmax}
\end{equation}
If the value of $r_m$ determined from equation (\ref{eq:rmax}) is
smaller than $R_{t,1}$, equation (\ref{eq:Eorbsep}) does not apply and
the galaxies never separate but undergo a quick merger.

The external tidal force on the larger galaxy can be written as
$F_{\rm tid} \equiv f \, r_m \sigma_1^2 / R_{t,1}^2$, where $f=1$
corresponds to the peak value responsible for the truncation.  The
galaxy pair remains bound at the distance of maximum separation if it
satisfies the {\it binding criterion}:
\begin{equation}
{F_{\rm grav} \over F_{\rm tid}} = {(1-\chi_1) \, (1+M_2/M_1) \over f} 
   \left({R_{t,1} \over r_m}\right)^3 > 1.
  \label{eq:mercond3}
\end{equation}

\section{Calculating the Tidal Field}
  \label{sec:app_tidal}

This appendix describes the procedure for calculating the tidal field
in the simulations, as well as some numerical tests.  The second
derivative of the potential is evaluated using tri-linear
interpolation from the eight nearest mesh points enclosing the center
of mass of the galaxy.  In order to remove the effects of small-scale
noise, I apply the Savitzky-Golay smoothing filters of the 4th order
(\citealt{NR}, \S 14.8, routine {\tt savgol}).  These filters replace
each data point $\Phi_i$ with a linear combination of itself and
several neighbors:
\begin{equation}
\Phi_i^* = \sum_{n=-N_L}^{N_R} \, c_n^{(0)} \, \Phi_{i+n},
  \label{eq:filter}
\end{equation}
where $N_L$ and $N_R$ are the number of points used ``to the left''
and ``to the right'', respectively.  The first and second derivatives
of the potential are obtained through similar expressions with
different coefficients $c_n^{(1)}$ and $c_n^{(2)}$.  The double
derivatives, $F_{xx}$, $F_{yy}$, $F_{zz}$, are calculated at the mesh
points $(i,j,k)$ as
\begin{equation}
F_{xx}(i,j,k) = {1 \over \Delta^2} \,
    \sum_{n=-N_L}^{N_R} \, c_n^{(2)} \, \Phi(i+n,j,k),
\end{equation}
where $\Delta$ is the mesh size.  The mixed derivatives, $F_{xy}$,
$F_{xz}$, and $F_{yz}$, are calculated using the smoothing in both
directions:
\begin{equation}
F_{xy}(i,j,k) = {1 \over \Delta^2} \,
    \sum_{n_1=-N_L}^{N_R} \sum_{n_2=-N_L}^{N_R} \, c_{n_1}^{(1)} \,
    c_{n_2}^{(1)} \, \Phi(i+n_1,j+n_2,k).
\end{equation}
The coefficients are chosen such that the 4th derivative (the
smoothing order) of the field is preserved.  I take $N_L = N_R = 4$,
which corresponds to the effective smoothing of the tidal field on the
scale of $250 - 500\, h^{-1}$ kpc.

The accuracy of the above smoothing has been tested against the
analytic NFW model \citep{NFW:97} with the parameters of the
Cluster II simulation:
\begin{equation}
\rho(r) = {\rho_0 \over (r/r_s) (1 + r/r_s)^2}.
\end{equation}
For the virial radius $R_{\rm vir} \approx 1.5\, h^{-1}$ Mpc and the
``concentration'' relation $R_{\rm vir} = 10\ r_s$, the scale radius
is $r_s = 0.15\, h^{-1}$ Mpc.  In the $512^3$ simulation the grid cell
size is $\Delta = 62.5\, h^{-1}$ kpc, so that $r_s \approx 2.5\,
\Delta$.  The exact tidal force in the NFW model is readily calculated
using equation (5) of \cite{GHO:99}.

Figure \ref{fig:testderiv} shows that our method reproduces well the
mixed derivative $F_{xy}$ down to 3 cell sizes from the center, and
somewhat worse the double derivative $F_{xx}$.  In the case of the
mixed derivative, the filter may seem to be less accurate than the
straightforward two-point finite difference.  But in real simulations
with numerical noise the smoothing filter becomes more effective and
removes small-scale artifacts.

\subsection{Test: The Pancake Simulation}

As another check of accuracy of the calculation of the tidal field, I
have run a test simulation of a one-dimensional Zeldovich's pancake.
The initial conditions are set up as described in \cite{AMR}.  The
wavelength of the sine wave perturbation equals the size of the
simulation box; the maximum displacement is in the center.  The
simulation is run until a caustic of infinite density forms.  From
Poisson's equation it follows that the only non-zero component of the
tidal tensor is $F_{xx} = 4\pi G \Delta\rho$, where $\Delta\rho$ is
the density perturbation.

I have run two simulations, with the high ($512^3$ grid) and low
($128^3$ grid) resolution.  The density profile in both simulations is
reproduced accurately.  The calculated tidal force matches the exact
solution in the beginning while the perturbation is small, but
deviates at later times.  The smoothing filter cannot resolve a very
narrow caustic.  As expected, the higher-resolution simulation departs
from the exact solution later than the lower-resolution one, at 70\%
vs 20\% of the total simulation time.  A similar effect of the
resolution loss in high density regions in real simulations is
emphasized again in the next section.

\subsection{Subtracting Galactic Self-Contribution}

Large halos may contribute significantly to the amplitude of the local
tidal field.  If the halo extends over more than a couple of cells, it
will contaminate the calculation of the {\it external} tidal force.
It is therefore necessary to correct for the galactic
self-contribution.

For this purpose, I have implemented a potential solver for an
isolated galaxy.  On a rectangular grid of $16^3$ cells, the density
of the galactic particles is assigned using the CIC method.
Then, the potential is obtained by the FFT method with isolated
boundary conditions.  Applying the same procedure for calculating the
tidal tensor as above, I evaluate and subtract the halo
self-contribution.

The FFT solver for Poisson's equation with isolated boundary
conditions is implemented following \cite{HE:81}.  The density grid is
doubled in all three directions and padded with zeros in the added
space.  Green's function is translated correctly to the new sections
in real space and then Fourier transformed and stored for the
potential calculation.  The trick with extending the grid is necessary
in order to avoid the multiple images which appear in the usual FFT
with periodic boundary conditions.  The resulting Green's function
$G_{\rm isol}(k)$ differs from the usual form, $G(k) = k^{-2}$, by
cutting off the high frequency tail.  Afterwards the algorithm
proceeds as usual: Fourier transforming the density field, multiplying
by $G_{\rm isol}(k)$, and transforming back to real space.

The isolated FFT procedure has been tested on the same $16^3$ grid,
using $2^{15}$ particles distributed in the central cell according to
the \cite{H:90} model:
\begin{equation}
\rho(r) = {M \over 2\pi} \, {a\over r} \, {1 \over (r+a)^3},
\end{equation}
with a tiny core radius, $a = 0.1\, \Delta$.  Essentially, this is a
point mass represented by particles.  The algorithm reproduces the
exact potential with very high accuracy, all the way to the edge of
the simulation box.  (For comparison, the periodic FFT produces large
errors on most of the grid.)  The force calculation is limited in the
center at $2\, \Delta$, as in any grid-based method, but behaves
perfectly well at larger radii.  The force calculated with the
periodic FFT shows much more noise.

Another test has been applied to make sure the procedure works in a
real simulation.  The $16^3$ density field was calculated using the
particles from a large galaxy, which were all located within a single
grid cell and resembled a smoothed point source.  Although with
increased scatter, the new method is still in a good agreement with
the exact potential and force.

Figure \ref{fig:testtidal} shows the self-contribution of a large
galaxy in the Cluster III simulation.  For the ease of comparison, I
plot the trace of the tidal tensor ($F_{xx}+F_{yy}+F_{zz}$) which is
generally proportional to the density of galactic particles, in accord
with Poisson's equation.  The smoothing filter misses the extreme
variations of the density, but otherwise the amplitude of the tidal
force is calculated correctly.

Also, Figure \ref{fig:testtidal} shows the scaling $(1+z)^3$, which
corresponds to a constant comoving density.  The halo density follows
this scaling quite closely indicating that the galaxy expands in real
space.  This numerical artifact is due to the loss of force resolution
within a grid cell.  The particles cannot collapse below 1 $\Delta$,
and therefore in comoving coordinates the galaxy continues to occupy
the same volume.



\begin{figure}
\plotone{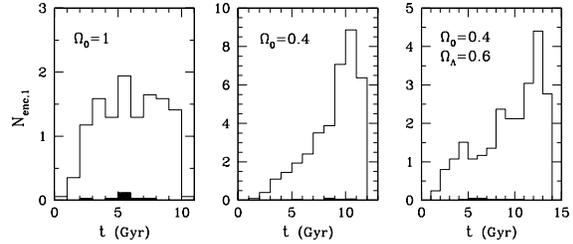}
\caption{Distribution of close encounters in time: Number of
  encounters with $R_{\rm enc} < 20$ kpc per galaxy per Gyr.  Filled
  histograms show the number of possible mergers.
  \label{fig:enctime}}
\end{figure}

\begin{figure}
\plotone{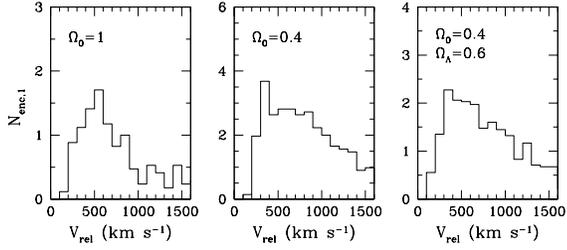}
\caption{Distribution of relative velocities: Number of encounters
  with $R_{\rm enc} < 20$ kpc per galaxy per 100 km s$^{-1}$.
  \label{fig:encvel}}
\end{figure}

\begin{figure}
\plotone{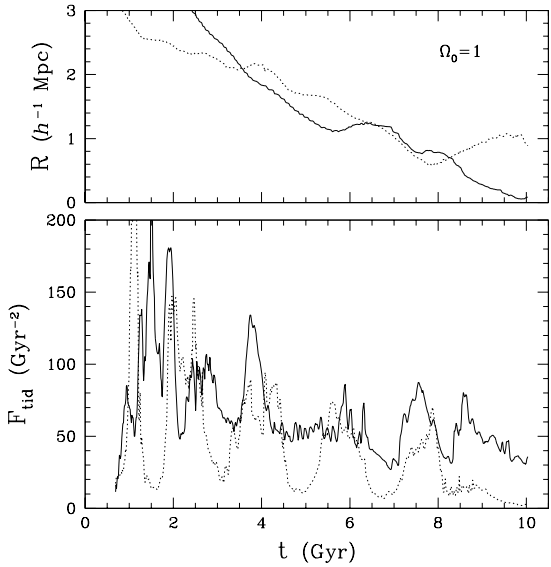}
\caption{Tidal field around the large (solid line) and dwarf (dots)
  galaxies in the $\Omega_0=1$ cluster
  (cf. Fig. \protect\ref{fig:traject_om1}).  Upper panel shows the
  distance from the final cluster center.
  \label{fig:tidal_om1}}
\end{figure}

\begin{figure}
\plotone{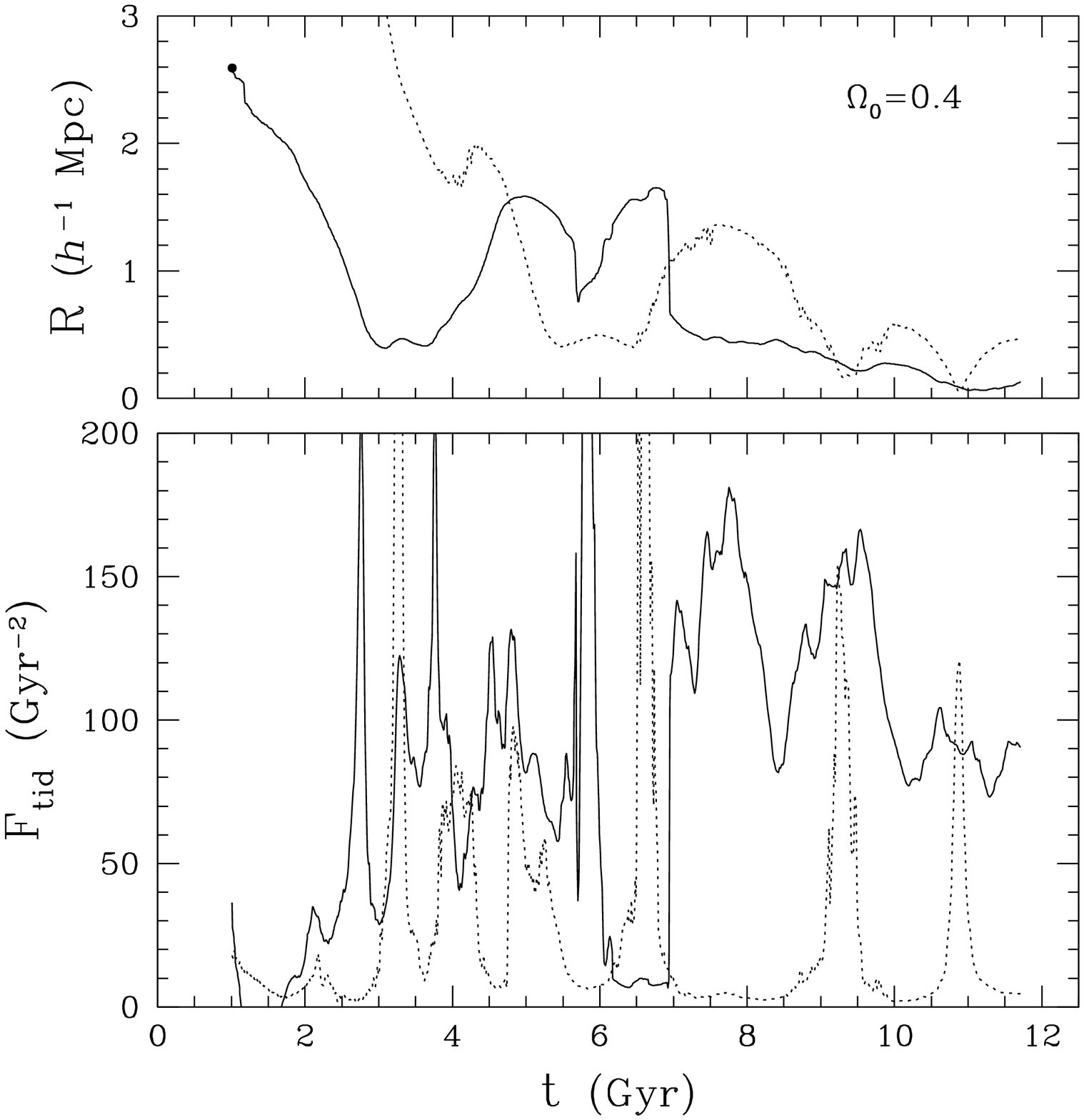}
\caption{Tidal field around the two galaxies in the $\Omega_0=0.4$ cluster
  (cf. Fig. \protect\ref{fig:traject_om.4}).
  Line notation as in Figure \protect\ref{fig:tidal_om1}.
  \label{fig:tidal_om.4}}
\end{figure}

\begin{figure}
\plotone{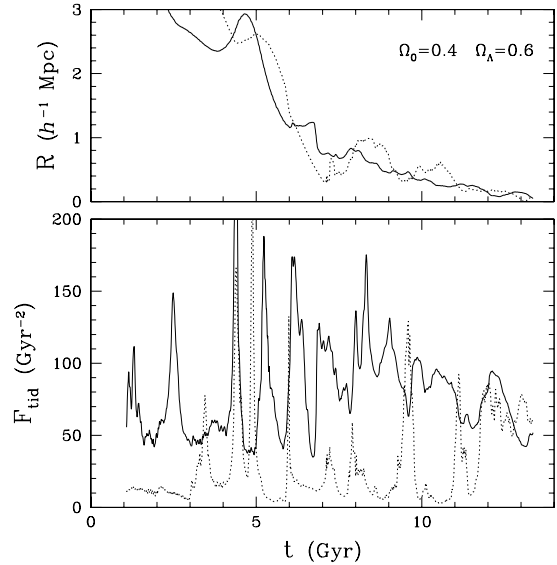}
\caption{Tidal field around the two galaxies in the $\Omega_0=0.4$,
  $\Omega_\Lambda=0.6$ cluster (cf. Fig. \protect\ref{fig:traject_om.4L}).
  Line notation as in Figure \protect\ref{fig:tidal_om1}.
  \label{fig:tidal_om.4L}}
\end{figure}

\clearpage

\begin{figure}
\plotone{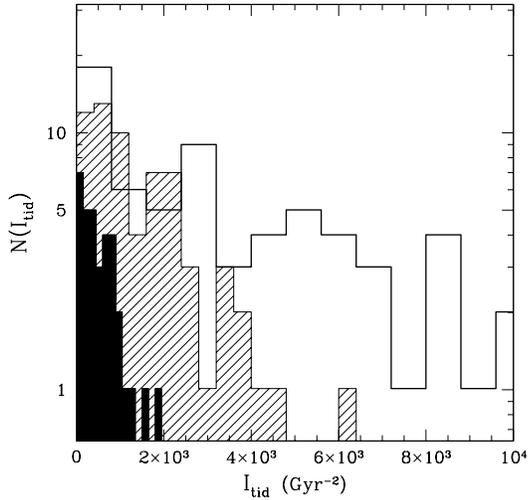}
\caption{The distribution of the tidal heating parameter, $I_{\rm
  tid}$ (eq. [\protect\ref{eq:ipar}]), for the identified galaxies in
  Cluster I (filled histogram), Cluster II (solid histogram), and
  Cluster III (hatched histogram).  
  \label{fig:tidalsum}}
\end{figure}

\begin{figure}
\plotone{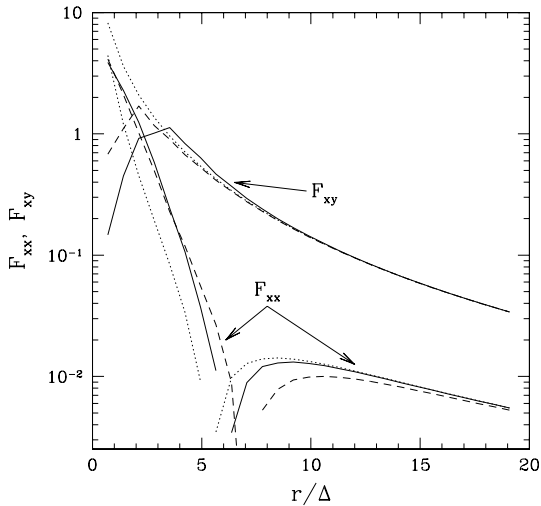}
\caption{The tidal tensor components in the NFW model (arbitrary units)
  versus radius in units
  of the cell size.  Dashed line is obtained with the direct two-point
  finite difference, solid line with our smoothing filter.  Dots show the
  analytic solution.  Note that $F_{xx}$ changes sign at
  $r \approx 5\, \Delta$; the inner curve shows the absolute value.
  \label{fig:testderiv}}
\end{figure}

\begin{figure}
\plotone{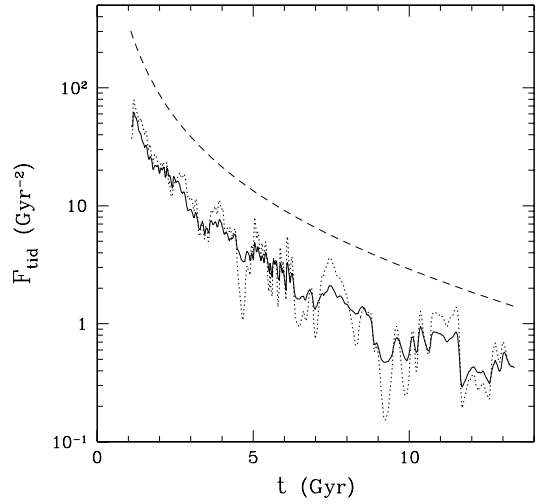}
\caption{Galaxy self-contribution to the trace of the tidal tensor
  (solid line) compared to the density of galactic particles ($4\pi G\rho$;
  dots) of halo \#4 in the Cluster III simulation.  Dashes show the
  $(1+z)^3$ scaling for a fixed comoving density (arbitrary
  normalization).
  \label{fig:testtidal}}
\end{figure}

\clearpage


\begin{deluxetable}{lccccc}
\tablecaption{Cosmological models \label{tab:cosmo}}
\tablecolumns{6}
\tablehead{\colhead{Cluster} & \colhead{$\Omega_0$} &
           \colhead{$\Omega_\Lambda$} & \colhead{$h$} & \colhead{$\sigma_8$}}
\tablewidth{0pt}
\startdata
I        &   1   &   0   &  0.65  &  0.5  \\
II       &  0.4  &   0   &  0.65  &  1.0  \\
III~~~~  &  0.4  &  0.6  &  0.65  &  1.0
\enddata
\end{deluxetable}

\begin{deluxetable}{lccccc}
\tablecaption{Parameters of the simulated clusters \label{tab:clusters}}
\tablecolumns{6}
\tablehead{\colhead{} & \colhead{$R_{\rm vir}$} & \colhead{$M_{\rm vir}$} &
           \colhead{$\sigma_{\rm los}$} & \colhead{$b/a$} & \colhead{$c/a$} \\
           \colhead{Cluster} & \colhead{($h^{-1}$ Mpc)} &
           \colhead{($M_{\sun}$)} & \colhead{(km s$^{-1}$)}}
\tablewidth{0pt}
\startdata
I        &  1.03  &  $3.9 \times 10^{14}$  &  656  &  0.70  &  0.58  \\
II       &  1.47  &  $4.5 \times 10^{14}$  &  664  &  0.85  &  0.71  \\
III~~~~  &  1.42  &  $4.1 \times 10^{14}$  &  649  &  0.85  &  0.73
\enddata
\end{deluxetable}

\begin{deluxetable}{lccccc}
\tablecaption{Resolution test on the parameters of Cluster III
              \label{tab:clusters_res}}
\tablecolumns{6}
\tablehead{\colhead{} & \colhead{$R_{\rm vir}$} & \colhead{$M_{\rm vir}$} &
           \colhead{$\sigma_{\rm los}$} & \colhead{$b/a$} & \colhead{$c/a$} \\
           \colhead{$N_{\rm grid}$} & \colhead{($h^{-1}$ Mpc)} &
           \colhead{($M_{\sun}$)} & \colhead{(km s$^{-1}$)}}
\tablewidth{0pt}
\startdata
$512^3$     &  1.42  &  $4.1 \times 10^{14}$  &  649  &  0.85  &  0.73  \\
$256^3$     &  1.42  &  $4.1 \times 10^{14}$  &  632  &  0.84  &  0.70  \\
$128^3$~~~~ &  1.43  &  $4.1 \times 10^{14}$  &  546  &  0.76  &  0.61
\enddata
\end{deluxetable}

\begin{deluxetable}{lcccc}
\tablecaption{Parameters of galaxy distribution\label{tab:gal}}
\tablecolumns{5}
\tablehead{\colhead{Cluster}              & \colhead{$N_g$}    &
           \colhead{$r_0$ ($h^{-1}$ kpc)} & \colhead{$\gamma$} &
           \colhead{${\cal C}$} \\
           \hline & & \colhead{All galaxies}}
\tablewidth{0pt}
\startdata
I        &  34  &   11  &  2.4  &  68   \\
II       &  69  &   70  &  4.2  &  510  \\
III~~~~  &  65  &   22  &  3.4  &  1600 \\
\cutinhead{Galaxies with $r > \Delta = 62.5\, h^{-1}$ kpc}
I        &  33  &   66  &  2.9  &  15  \\
II       &  61  &  110  &  4.5  &  230 \\
III      &  43  &   59  &  3.3  &  100
\enddata
\end{deluxetable}

\begin{deluxetable}{lccccc}
\tablecaption{Number of mergers and close encounters\label{tab:enc}}
\tablecolumns{6}
\tablehead{& \colhead{$R_{\rm enc} < 5$ kpc} & \colhead{10 kpc}
           & \colhead{15 kpc} & \colhead{20 kpc} & \colhead{Total}}
\tablewidth{0pt}
\startdata
\cutinhead{Cluster I}
$N_{\rm enc}$         &   32 &   97 &  139 &  211 &  498 \\
$N_{\rm mer}$         &    0 &    5 &    8 &    8 &   10 \\
$N_{\rm enc,1}$       &  1.9 &  5.7 &  8.2 &   12 &   29 \\
$N_{\rm mer,1}$       &    0 & 0.15 & 0.24 & 0.24 & 0.29 \\
$\tilde{N}_{\rm enc}$ &   32 &  130 &  241 &  381 &  890 \\
$N_{\rm enc}$ (test)  &    3 &   23 &   43 &   80 &  213 \\
$N_{\rm mer}$ (test)  &    0 &    0 &    0 &    0 &    1 \\ 
\cutinhead{Cluster II}
$N_{\rm enc}$         &  435 &  866 & 1128 & 1280 & 2576 \\
$N_{\rm mer}$         &    0 &    9 &   18 &   20 &   20 \\
$N_{\rm enc,1}$       &   13 &   26 &   33 &   38 &   76 \\
$N_{\rm mer,1}$       &    0 & 0.13 & 0.26 & 0.29 & 0.29 \\
$\tilde{N}_{\rm enc}$ &  435 & 1404 & 2596 & 3590 & 6631 \\
$N_{\rm enc}$ (test)  &   19 &   62 &  143 &  273 &  902 \\
$N_{\rm mer}$ (test)  &    0 &    0 &    0 &    0 &    0 \\
\cutinhead{Cluster III}
$N_{\rm enc}$         &   95 &  314 &  575 &  782 & 1877 \\
$N_{\rm mer}$         &    4 &    7 &   10 &   11 &   14 \\
$N_{\rm enc,1}$       &  3.0 &  9.8 &   18 &   24 &   59 \\
$N_{\rm mer,1}$       & 0.06 & 0.11 & 0.15 & 0.17 & 0.22 \\
$\tilde{N}_{\rm enc}$ &  136 &  525 &  962 & 1360 & 3113 \\
$N_{\rm enc}$ (test)  &   13 &   59 &  129 &  197 &  614 \\
$N_{\rm mer}$ (test)  &    0 &    0 &    0 &    0 &    1
\enddata
\end{deluxetable}

\begin{deluxetable}{lccc}
\tablecaption{Relative velocities in close encounters\label{tab:relvel}}
\tablecolumns{4}
\tablehead{ & \colhead{$V_{\rm peak}$} & \colhead{$V_{\rm med}$} &
              \colhead{$V_{\rm mean}$} \\
            \colhead{Cluster} & \colhead{(km s$^{-1}$)} &
            \colhead{(km s$^{-1}$)} & \colhead{(km s$^{-1}$)}}
\tablewidth{0pt}
\startdata
I       & 550 & 686 &  936 \\
II      & 350 & 872 & 1196 \\
III~~~~ & 350 & 817 & 1097
\enddata
\end{deluxetable}

\begin{deluxetable}{lccc}
\tablecaption{Galaxy Kinematics\label{tab:disp}}
\tablecolumns{4}
\tablehead{\colhead{Region} & \colhead{$< r$} & \colhead{$N_g(r)$}
           & \colhead{$\sigma_{g,\rm los}$}\\
	   & \colhead{($h^{-1}$ Mpc)} & & \colhead{(km s$^{-1}$)}}
\tablewidth{0pt}
\startdata
\cutinhead{Cluster I}
Inner         & 0.21  & 11 & 373 \\
Middle        & 0.65  & 11 & 324 \\
$R_{\rm vir}$ & 1.03  & 12 & 538 \\
\cutinhead{Cluster II}
Inner         & 0.11  & 23 & 117 \\
Middle        & 0.34  & 23 & 433 \\
$R_{\rm vir}$ & 1.47  & 23 & 469 \\
\cutinhead{Cluster III}
Inner         & 0.063 & 22 & 131 \\
Middle        & 0.5   & 22 & 324 \\
$R_{\rm vir}$ & 1.42  & 21 & 517
\enddata
\end{deluxetable}

\end{document}